\documentstyle[onecolumn,graphics]{mn}
\newcommand{\etal}{{et al.}~}

\newcommand{\de}{\delta}
\newcommand{\f}{\frac}
\newcommand{\lam}{\lambda}
\newcommand{\Lam}{\Lambda}
\newcommand{\mm}{{\mit m}}
\newcommand{\p}{\partial}

\newcommand{\eps}{\epsilon}
\newcommand{\Om}{\Omega}
\newcommand{\w}{\omega}
\newcommand{\s}{\sigma}
\newcommand{\al}{\alpha}
\newcommand{\fde}{\tilde{\delta}}
\newcommand{\fpsi}{\tilde{\psi}}

\newcommand{\fW}{\widetilde{W}}
\newcommand{\bfx}{{\bf x}}
\newcommand{\bfr}{{\bf r}}

\newcommand{\bfv}{{\bf v}}
\newcommand{\bfq}{{\bf q}}

\newcommand{\bfp}{{\bf p}}
\newcommand{\bfu}{{\bf u}}
\newcommand{\bfL}{{\bf L}}
\newcommand{\bfS}{{\bf S}}

\newcommand{\calA}{{\cal A}}
\newcommand{\calB}{{\cal B}}
\newcommand{\calW}{{\cal W}}
\newcommand{\calD}{{\cal D}}
\newcommand{\calE}{{\cal E}}
\newcommand{\calF}{{\cal F}}

\newcommand{\calJ}{{\cal J}}
\newcommand{\calL}{{\cal L}}

\newcommand{\calP}{{\cal P}}
\newcommand{\bc}{\begin{center}}
\newcommand{\be}{\begin{equation}}
\newcommand{\ee}{\end{equation}}
\newcommand{\ec}{\end{center}}
\newcommand{\lan}{\langle}
\newcommand{\ran}{\rangle}
\title[Linear evolution of the tidal angular momentum]
{Evolution of the angular momentum
of protogalaxies from tidal torques: Zel'dovich approximation} 

\author[P. Catelan and T. Theuns]
{Paolo Catelan and Tom Theuns\\
Department of Physics, Astrophysics, University of Oxford, Keble Road, 
Oxford OX1 3RH, UK \\} 

\begin{document}

\maketitle

\begin{abstract}
The growth of the angular momentum $\bfL$ of protogalaxies induced by
tidal torques is reconsidered.  We adopt White's formalism and study
the evolution of $\bfL$ in Lagrangian coordinates; the motion of the
fluid elements is described by the Zel'dovich approximation. We obtain
a general expression for the ensemble expectation value of the square
of $\bfL$ in terms of the first and second invariant of the inertia
tensor of the Lagrangian volume $\Gamma$ enclosing the protoobject's
collapsing mass. We then specialize the formalism to the particular
case in which $\Gamma$ is centred on a peak of the smoothed Gaussian
density field and approximated by an isodensity ellipsoid. The result
is the appropriate analytical estimate for the rms angular momentum of
peaks to be compared against simulations that make use of the
Hoffman-Ribak algorithm to set up a constrained density field that
contains a peak with given shape. Extending the work of Heavens \&
Peacock, we calculate the {\it joint} probability distribution
function for several spin parameters and peak mass $M$ using the
distribution of peak shapes, for different initial power spectra. The
probability distribution for the rms final angular momentum $\langle
\bfL_f^2\rangle^{1/2}$ on the scales corresponding to common bright
galaxies, $M\approx 10^{11} M_{\odot}$, is centred on a value of
$\approx 10^{67}\, {\rm kg}\,{\rm m}^2\,{\rm s}^{-1}$, for any
cosmologically relevant power spectrum, in line with previous
theoretical and observational estimates for $L_f$. Other astrophysical
consequences are discussed. In particular, we find that typical values
$\lan \lambda^2\ran^{1/2}\approx 0.1$ of the dimensionless spin
parameter for peaks smoothed on galactic scales and of height $\nu\sim
1$, usually associated with late type galaxies, may be recovered in
the framework of the Gaussian peak formalism. This partially relaxes
the importance attributed to dissipative processes in generating such
high values of centrifugal support for spiral galaxies. In addition,
the values of the specific angular momentum versus mass -- as deduced
from observations of rotational velocities and photometric radii of
spiral galaxies -- are well fitted by our theoretical isoprobability
contours. In contrast, the observed lower values for the specific
angular momentum for ellipticals of the same mass cannot be accounted
for within our linear regime investigation, highlighting the
importance of strongly non-linear phenomena to explain the spin of
such objects.
\end{abstract}


\section{Introduction}
It has been argued that tidal coupling between the inhomogeneities in
the primordial matter distribution, in the context of a gravitational
hierarchical scenario, may explain the acquisition of the angular
momentum by a protogalaxy.  This idea, originally due to Hoyle (1949)
and applied by Sciama (1955), has been first thoroughly examined by
Peebles (1969), who demonstrated that the tidal spin growth of the
matter contained in a spherical (Eulerian) volume is proportional to
$t^{5/3}$ in an Einstein--de Sitter universe ($t$ is the standard
cosmic time).  Specifically, Peebles' analysis is based on a {\it
second-order} perturbative description, since a spherical volume does
not gain angular momentum from tidal torques in linear approximation,
as pointed out by Zel'dovich and reported by Doroshkevich (1970).  An
important result of Doroshkevich's paper is that a {\it generic
nonspherical} volume enclosing the protogalaxy acquires tidal angular
momentum proportionally to the cosmic time $t$ during the linear
regime.  This theoretical prediction has been confirmed by the
$N$--body simulations of White (1984). In addition, White showed that
the second-order growth described by Peebles is due to convective
motion of matter across the surface of the initial volume $\Gamma$
containing the protoobject.  An important point of Peebles and White's
theoretical analyses is that they describe the tidal torques acting on
a volume centred on a {\it random} point in the smoothed density
field, which does not necessarily enclose a bound protogalaxy. In
contrast, galaxies are expected to form around (high) peaks on
relevant scales in the density field.  This idea, in embryo in
Doroshkevich (1970), developed into the biased galaxy formation
scenario, where only density maxima above a given threshold (peaks) of
the initial Gaussian density field can eventually form galaxies
(Kaiser 1984; Politzer \& Wise 1984; Peacock \& Heavens 1985; Bardeen
\etal 1986).

The acquisition of angular momentum due to tidal torques by Gaussian
high--density peaks has been recently analyzed, amongst others, by
Hoffman (1986; 1988), Heavens
\& Peacock (1988), Ryden (1988), Quinn \& Binney (1992) and Eisenstein
\& Loeb (1995). Comparisons with $N$--body simulations are
displayed in Efstathiou \& Jones (1979), Barnes \& Efstathiou (1987)
and Warren \etal (1992).

In this paper, we re-examine the growth during the linear regime of
the angular momentum $\bfL$ of protogalaxies induced by tidal
couplings with the surrounding matter inhomogeneities. The layout of
this article is the following. In the next section we briefly review
White's formalism describing the linear evolution of tidal galactic
spin. The motion of the mass fluid elements is described by the
Zel'dovich approximation: the invariance of the angular momentum with
respect to the Eulerian and Lagrangian description is stressed.  Next,
we derive a general (but approximate) expression for the ensemble
expectation value of the square of $\bfL$, $\lan \bfL^2\ran$, in terms
of the first and second invariant of the inertia tensor of the
Lagrangian volume $\Gamma$. We then specialize our formalism to the
particular case in which the Lagrangian volume is centred on a peak of
the underlying smoothed Gaussian density field and approximated by an
isodensity profile ellipsoid: in this case, we obtain the correct
constrained ensemble average for these objects with preselected
inertia tensor. The result is the appropriate analytical estimate for
the rms angular momentum of peaks to be compared against simulations
that make use of the Hoffman-Ribak algorithm to set up a constrained
density field that contains a peak with given shape (Hoffman \& Ribak
1991; van de Weygaert \& Bertschinger 1996).  Extending the work of
Heavens \& Peacock, we calculate the {\it joint} probability
distribution function for several spin parameters (angular momentum,
specific angular momentum, angular momentum in units of $M^{5/3}$) and
peak mass $M$ using the distribution of peak shapes (Bardeen et
al. 1986), for different density power spectra.  Finally, we discuss
astrophysical implications in the last section. Technical details are
given in appendices.
 
\section{Angular momentum}
We carry out our analysis of the evolution of the angular momentum in
three steps: first, we review White's method for obtaining an
expression for ${\bf L}$ involving the shape of the protoobject and
the distribution of the surrounding matter. Next, we simplify the
expression for $\bfL$ by performing the ensemble average in
subsection~2.2. Finally, we specialize the formalism to the case where
the object is centered on a peak of the (Gaussian) underlying density
distribution.

\subsection{Dynamical description}
Let us assume that at the epoch of structure formation the matter may
be described on the relevant scales as a Newtonian collisionless cold
fluid (dust) embedded in an expanding Friedmann universe with
arbitrary density parameter $\Om$ (but, for simplicity, vanishing
cosmological constant). We indicate the comoving Eulerian spatial
coordinate by $\bfx$. The physical distance is $\bfr = a(t)\bfx$,
with $a(t)$ the expansion scale factor. In the Einstein--de Sitter
model, $a(t)\propto t^{2/3}$.

The angular momentum $\bfL$ of the matter contained at time $t$ in a
volume $V$ of the Eulerian ${\bfx}$-space is
\be
\bfL(t)=\int_{a^3V}d\bfr\,\rho\bfr\times\bfv=
\rho_b a^4\int_Vd\bfx\,(1+\de)\bfx\times\bfu\;.
\label{eq:Lorig}
\ee
Here, $\rho=\rho_b(1+\de)$ denotes the matter density field, $\rho_b$
the background mean density, $\de$ the density fluctuation field;
$\bfv=d\bfr/dt$ indicates the velocity field and $\bfu=a\,d\bfx/dt$ is
the so called ``peculiar'' velocity (see, e.g., Peebles 1980). The
origin of the Cartesian coordinate system is assumed to coincide with
the centre of mass. Since we are interested in the intrinsic angular
motion, we disregard the centre--of--mass motion.

It is important to note that the previous integral over the
$Eulerian$ volume $V$ may equally be written as an integral
over the corresponding $Lagrangian$ volume $\Gamma$:
\be
\bfL(t)=\eta_0\,
a^2\int_{\Gamma}d\bfq\;(\bfq+\bfS)\times\f{d\bfS}{dt}\;,
\label{eq:Lt}
\ee
where $\rho_b a^3=\rho_0 a_0^3 \equiv \eta_0$ in the matter dominated
era. Equation~(\ref{eq:Lt}) can be obtained from equation~(\ref{eq:Lorig})
using the mapping
\be
\bfx(\bfq, t)=\bfq + \bfS(\bfq, t)\;,
\label{eq:lag}
\ee
from Lagrangian coordinates ${\bf q}$ to Eulerian coordinates ${\bf
x}$, where ${\bf S}$ is the displacement vector. The determinant $J$
of the Jacobian of the mapping $\bfq\rightarrow
\bfx(\bfq, t)$, $d\bfx \,J^{-1}=d\bfq$, is related to the density fluctuation through the
continuity equation:
\be
1+\de[\bfx(\bfq, t), t] = J(\bfq, t)^{-1}\;.
\label{eq:cont}
\ee
Before the occurrence of shell crossing (caustic formation process) $J$
is non vanishing (see, e.g., Shandarin \& Zel'dovich 1989).

The expression for $\bfL(t)$ in equation~(\ref{eq:Lt}) allows us to
apply the $Lagrangian$ theory directly: we stress that it is an
$exact$ relation. Consequently, when applying perturbation theory to
equation~(\ref{eq:lag}), perturbative corrections to $\bfS$ (Bouchet
\etal 1992; Buchert 1994; Catelan 1995 and references therein)
provide perturbative corrections to $\bfL$ (Catelan \& Theuns
1996). Furthermore, equation~(\ref{eq:Lt}) shows that $\bfL(t)$
depends on the shape of the Lagrangian volume $\Gamma$, which
encloses, by definition, all the fluid elements that eventually form
the galaxy. The boundary of a given condensation region $\Gamma$
depends on the particular realisation of the density field and in
general it may be extremely fuzzy.

The linear Lagrangian theory corresponds to the Zel'dovich approximation,
\be
\bfS(\bfq, t)\approx \bfS^{(1)}(\bfq, t) = D(t)\nabla\psi_1(\bfq),
\label{eq:Zel}
\ee
which coincides with the linear Eulerian approximation to first order
in the spatial positions, and gives a qualitative correct
quasi--nonlinear solution by extrapolating the straight trajectories
of equation (\ref{eq:Zel}) beyond the range of strict validity
(Zel'dovich 1970a, b; Shandarin \& Zel'dovich 1989).  The temporal
function $D(t)$ describes the growing mode of the density
perturbations: in the Einstein--de Sitter model $D(t)\propto a(t)$.
We neglect the decaying mode.

The potential $\psi_1(\bfq)$ is the linear gravitational potential
(see the discussion in Kofman 1991): if $\fpsi_1(\bfp)\equiv\int
d\bfq\,\psi_1(\bfq)\,{\rm e}^{-i\,\bfp\cdot\bfq}$ is the Fourier
transform of the gravitational potential, then the Fourier transform
$\fde_1(\bfp)$ of the linear density field $\de^{(1)}(\bfq,
t)=D(t)\de_1(\bfq)$ is related to $\fpsi_1(\bfp)$ via the Poisson
equation, $\fde_1(\bfp)=p^2\,\fpsi_1(\bfp)$, where $p\equiv|\bfp|$,
and $\bfp$ is the comoving Lagrangian wavevector.

In the Zel'dovich approximation, the exact equation (\ref{eq:Lt})
simplifies to:
\be
\bfL(t)\approx \bfL^{(1)}(t)=\eta_0\, a^2\int_{\Gamma}d\bfq\,
\Big(\bfq+\bfS^{(1)}\Big)\times\f{d\bfS^{(1)}}{dt}
=\eta_0\, a^2
\f{dD}{dt}\int_{\Gamma}d\bfq\,\bfq\times\nabla\psi_1(\bfq)\;.
\label{eq:L-Zel}
\ee

As in White (1984), if it is assumed that $\psi_1(\bfq)$ is adequately
represented in the generic volume $\Gamma$ by the first three terms of
the Taylor series about the origin $\bfq={\bf 0}$, then each component
$L^{(1)}_{\al}(t)$ may be written in a compact form as (White 1984)
\be
L^{(1)}_{\al}(t)=a(t)^2\dot{D}(t)\,\eps_{\al\beta\gamma}\,
\calD^{(1)}_{\beta\s}\,\calJ_{\s\gamma}\;,
\label{eq:L1_White}
\ee
once the deformation tensor at the origin $\bfq={\bf 0}$
\be
\calD^{(1)}_{\beta\s}\equiv\calD^{(1)}_{\beta\s}({\bf 0})=
\p_{\beta}\p_{\s}\psi_1({\bf 0})\;,
\label{eq:defor}
\ee
and the inertia tensor of the mass contained in the
volume $\Gamma$
\be
\calJ_{\s\gamma}\equiv\eta_0\int_{\Gamma}d\bfq\,q_{\s}\,q_{\gamma}\;,
\label{eq:inertia}
\ee
are introduced. The dot in equation~(\ref{eq:L1_White}) represents the
operator $d/dt$, the symbol $\p_{\al}\equiv \p/\p q_{\al}$ and
$\eps_{\al\beta\gamma}$ is the completely antisymmetric Levi--Civita
tensor, $\eps_{123}=1.$ Summation over repeated Greek indices is
understood.

Equation~(\ref{eq:L1_White}) shows that the linear angular momentum
$\bfL^{(1)}$ is in general non-zero because the principal axes of the
inertia tensor $\calJ_{\al\beta}$, which depend only on the
(irregular) shape of the volume $\Gamma$, are not aligned with the
principal axes of the deformation tensor $\calD^{(1)}_{\al\beta}$,
which depend on the location of neighbour matter fluctuations (see the
discussion in White 1984). This fact can be illustrated by writing
explicitly one component of $\bfL$ in the inertia tensor's
principal-axes system, e.g., $L_1^{(1)}\propto
\calD_{23}^{(1)}\,(\calJ_{33}-\calJ_{22})$: only the {\it off-diagonal~}
elements of the deformation tensor enter.

In addition, the time dependence of $\bfL^{(1)}$ is governed by the
function $a(t)^2\dot{D}(t)$, which simplifies to $\f{2}{3}t_0^{-2}t$
in the Einstein--de Sitter universe, as first noted by Doroshkevich
(1970) (the convention $a=D=(t/t_0)^{2/3}$ is adopted). Finally, we
emphasize the spherical case: if $\Gamma$ is a spherical Lagrangian
volume, then
$\calJ_{\s\gamma}=\f{4\pi}{15}\,\eta_0\,q^5\,\de_{\s\gamma}$ and
$L^{(1)}_{\al}\propto
\eps_{\al\beta\gamma}\,\calD^{(1)}_{\beta\gamma}=0$; the symbol
$\de_{\al\beta}$ indicates the Kronecker tensor. Consequently, the
matter initially contained in a spherical volume does not gain any
tidal spin during the linear regime (see the discussion in White
1984). Of course, as we will show in the next section, this also holds
for the variance, i.e., $\lan\bfL^{(1)2}_{sphere}\ran=0$.

Note that the angular momentum $L^{(1)}_{\al}$ may be written 
directly in terms of the gravitational tidal field (which coincides
with the shear field in the linear regime)
\be
\calE^{(1)}_{\beta\s}\equiv 
\calD^{(1)}_{\beta\s}-\f{1}{3}\,D^{-1}\,(\nabla\cdot\bfS^{(1)})\de_{\beta\s}
= (\p_{\beta}\p_{\s}-\f{1}{3}\de_{\beta\s}\nabla^2)\psi_1\;,
\ee
as 
\be
L^{(1)}_{\al}(t)=a(t)^2\dot{D}(t)\,\eps_{\al\beta\gamma}\,
\calE^{(1)}_{\beta\s}\,\calJ_{\s\gamma}\;,
\ee
since the isotropic part of the deformation tensor does not contribute
to the antisymmetric tensor product: this justifies the use of
$tidal\,$ angular momentum to indicate the angular momentum originated
by the coupling between the {\it traceless} part of the deformation
tensor and the inertia tensor.

Finally, one can express the initial deformation tensor
$\calD^{(1)}_{\al\beta}$ as an integral in Fourier space: 
\be
\calD^{(1)}_{\al\beta}=
-\int\f{d\bfp}{(2\pi)^3}\,p_{\al}\,p_{\beta}\,\fpsi_1(\bfp)\,\fW(pR)\;.
\label{eq:defilt}
\ee
Note that, here and hereafter, we are explicitly assuming that the
potential field $\psi_1$ is filtered on scale $R$ by means of the
window function $W_R(\bfq)$. The function $\fW(pR)$ is the Fourier
transform of the smoothing function $W_R(\bfq)$. The filtering of the
field $\psi_1$ on an appropriate scale reflects the restriction to the
linear evolution of the protogalaxy, since the non-linear coupling
between different modes is filtered out.

As far as one considers the evolution of a single collapsing region of
assigned enclosing volume $\Gamma$ and velocity field, then, to
calculate its final angular momentum acquired via tidal couplings with
the neighbour matter distribution, it is enough to supplement the
equation~(\ref{eq:L1_White}) with the time $t_M$ at which the
perturbation stops expanding and starts to collapse. After $t_{M}$ the
angular momentum essentially stops growing, becoming less sensitive to
tidal couplings (Peebles 1969). While for a general aspherical
perturbation the epoch $t_{M}$ is not well defined (see Eisenstein \&
Loeb 1995 for the case of an ellipsoid), if one restricts study to a
spherical top--hat region with uniform overdensity
$\bar{\de}=\f{3}{20}(6\pi t/t_{M})^{2/3}$, then the time of maximum
expansion is reached when $
\bar{\de}=\f{3}{20}(6\pi)^{2/3}\,
$ (see Peebles 1980).

However, one is interested in calculating the tidal galactic spin
averaged over an $ensemble$ of realizations of the gravitational
potential random field $\psi_1$, rather than in the evolution of the
single perturbation.  This is particularly important in order to
compare the results from $N$--body simulations against theoretical
ones. This programme is carried out in the next section.

\subsection{Ensemble average}
\label{subsect:ensem}
We simplify the previous results by considering the expectation value
of the square of $\bfL$, $\lan \bfL^2\ran_{\psi}$. In linear theory
this is quite simple to do because
$\lan\bfL^2\ran_{\psi}=\lan\bfL^{(1)2}\ran_{\psi}$. Here and
hereafter, we indicate the $linear$ angular momentum simply by $\bfL$,
and the initial gravitational potential by $\psi\equiv\psi_1$.

From the expression~(\ref{eq:L1_White}) of $L_{\al}$, we obtain
\be
\lan\bfL^2\ran_{\psi}=
a^4\dot{D}^2\,\eps_{\al\beta\gamma}\,\eps_{\al\beta'\gamma'}\,
\lan\,\calJ_{\s\gamma}\,\calJ_{\s'\gamma'}\,\calD_{\beta\s}\calD_{\beta'\s'}\ran_{\psi}\approx
a^4\dot{D}^2\,\eps_{\al\beta\gamma}\,\eps_{\al\beta'\gamma'}\,
\calJ_{\s\gamma}\,\calJ_{\s'\gamma'}\,\lan\,\calD_{\beta\s}\calD_{\beta'\s'}\ran_{\psi}
\;.
\label{eq:L2ensem}
\ee
Note that the correct procedure to compute $\lan\bfL^2\ran_{\psi}$
would require complete understanding of how the different realizations
of $\psi$ stochastically influence the boundary of the generic volume
$\Gamma$, which in turn determines the inertia tensor $\calJ$. This is
a major unsolved problem. To bypass it, we have assumed in the last
step of equation~(\ref{eq:L2ensem}) that $\calJ$ is uncorrelated with
$\psi$, thereby simplifying the calculation considerably; the price to
pay is a loss of accuracy. This approximation is equivalent to
preselecting a volume with given inertia tensor and independently
assigning the realizations of the field, or, from a different point of
view, the last step of equation~(\ref{eq:L2ensem}) would be strictly
{\it correct} if the random field $\psi$ were to be restricted to
those realizations of the ensemble which are compatible with the
preselected inertia tensor and one would restrict study to those
objects only. For example, the Hoffman-Ribak algorithm (Hoffman \&
Ribak 1991; van de Weygaert \& Bertschinger 1996; Sheth 1995) can be
used to set up constrained realizations of the (Gaussian or
non-Gaussian) random density field containing objects with a given
inertia tensor.

In the context of the Gaussian peak formalism (section~2.3 and
Appendix~B), it is possible to compute exactly the (small) factor by
which the neglect of correlation between inertia tensor and the
gravitational potential in equation~(\ref{eq:L2ensem}) leads to an
overestimate of the rms angular momentum. Nevertheless, we prefer to
maintain the present procedure in this section, both for its
simplicity and because it is applied when calculating perturbative
corrections to the linear tidal angular momentum, whose distribution
deviates non-trivially from the Gaussian one during the mildly
non-linear regime (Catelan \& Theuns 1996; see also the last paragraph
in Appendix~B).

The ensemble average in equation~(13) can be reduced to an integral
over the initial power spectrum $P_{\psi}(k)$ using
equation~(\ref{eq:defilt}):
\be
\lan\calD_{\beta\s}\calD_{\beta'\s'}\ran_{\psi}=
\int\f{d\bfp}{(2\pi)^3}\,p_{\beta}\,p_{\s}\,p_{\beta'}\,p_{\s'}\,
P_{\psi}(p)\,\fW(pR)^2\;,
\label{eq:dd}
\ee
where
$
\lan\fpsi(\bfp)\fpsi(\bfp')\ran_{\psi}
\equiv (2\pi)^3\de_D(\bfp+\bfp')P_{\psi}(p)
$: the presence of the Dirac function is necessary to guarantee that
the density power spectrum has the correct units of volume and the
conservation of momentum in Fourier space.

Observing that in the integral~(\ref{eq:dd}) the function
$P_{\psi}(p)\,\fW(pR)^2$ depends only on the modulus $p$ of the
wavevector $\bfp$, and that the volume filtered by the window function
$\fW(pR)$ is spherically symmetric, applying the general
relation
\be
\int_{sphere}d\bfp\,p_{\al}\,p_{\beta}\,p_{\gamma}\,p_{\de}\,F(|\bfp|)=
\f{4\pi}{15}
(\de_{\al\beta}\,\de_{\gamma\de}+\de_{\al\gamma}\,\de_{\beta\de}+
\de_{\al\,\de}\,\de_{\beta\,\gamma})\int dp\,p^6\,F(p)\;,
\ee
which holds for any function $F(p)$ depending only on the radius
$p$, we finally obtain (see Appendix A for a detailed derivation)
the particularly simple result
\be
\lan\bfL^2\ran_{\psi}=\f{2}{15}a^4\dot{D}^2(\mu_1^2-3\mu_2)\,\s_0(R)^2\;.
\label{eq:Lensem}
\ee
This result holds for any power spectrum $P_{\psi}(p)$ and is
independent of the statistics of the underlying field. The quantity
$\s_0(R)^2$ is the mass variance on the scale $R$, i.e.  $
\s_0(R)^2\equiv (2\pi^2)^{-1}\int_0^{\infty} dp\,p^6\,P_{\psi}(p)\fW(pR)^2\;.
$

The general expression~(\ref{eq:Lensem}) is $independent\,$ of the
details of the shape of the boundary surface of the volume $\Gamma$:
it depends only on the quantities $\mu_1$ and $\mu_2$, which are
respectively the first and the second invariant of the inertia tensor
$\calJ_{\al\beta}$. The latter are defined in terms of the eigenvalues
$\iota_1$, $\iota_2$ and $\iota_3$ of the inertia tensor,
\be
\mu_1\equiv\iota_1+\iota_2+\iota_3\;,
\ee
\be
\mu_2\equiv\iota_1\iota_2+\iota_1\iota_3+\iota_2\iota_3\;.
\ee
(The third invariant is the determinant, $\iota_1\iota_2\iota_3$.)
We stress that the particular combination $\mu_1^2-3\mu_2$ is null for
a sphere, since $\iota_1=\iota_2=\iota_3$ for a spherical volume
$\Gamma$, whereas for a non-spherical volume, $\mu_1^2-3\mu_2>0$.

So far we have described the tidal torques acting on a volume centred
on a $random$ point in the smoothed density field, which does not
necessarily enclose a bound protogalaxy. However, galaxies are
expected to form around (high) peaks on relevant scales in the density
field: in the biased galaxy formation scenario, only the density
maxima above a given threshold (peaks) of the initial Gaussian density
field can eventually form galaxies (Kaiser 1984; Politzer \& Wise
1984; Peacock \& Heavens 1985; Bardeen \etal 1986). In the next
section, we explicitly assume the volume $\Gamma$ to be centred on a
peak of the underlying smoothed density field. In this case it is
possible to take into account {\it exactly} the correlation between
the inertia tensor of the matter contained in $\Gamma$ and the
potential (and hence also density) field. In the following we will
still use the notation $\lan\bfL^2\ran_\psi$ to denote the average
over the ensemble of realizations of $\psi$ which are compatible with
peaks with preselected inertia tensor,
$\lan\bfL^2\ran_{\psi|\calJ}$. The latter corresponds to the average
over the (unconstrained) realizations of off-diagonal tidal field (see
Appendix~B).

\subsection{Tidal torques and density peaks}
In this subsection, we specialize the ensemble average
$\lan\bfL^2\ran_{\psi}$ to the case in which the volume $\Gamma$ is
centered on a peak of the Gaussian density field. Consequently, the
combination of eigenvalues $\mu_1^2-3\mu_2$ will depend on the
parameters characterizing the peak's shape. The main difficulty to
overcome, however, remains that of estimating which fraction of the
matter surrounding the primordial peak will eventually collapse onto
the peak itself. In other words, the main problem is to estimate the
shape of the surface boundary of $\Gamma$. This is of course a very
difficult and unsolved problem.

Previous analyses of the topology of the constant-density profiles in
the neighborhood of the peaks of the Gaussian field $\de$ showed that
the isodensity surfaces $\de_c=\nu_c\s_0$, where $\nu_c$ is a critical
threshold in units of $\s_0$, are simply connected and approximately
ellipsoidal, at least for sufficiently high density peaks
(Doroshkevich 1970; Bardeen \etal 1986; Couchman 1987).  Hence, we
assume that the surface boundary of the volume $\Gamma$ is determined
by the criterion that all the matter above the threshold $\nu_c=0$
will collapse and form the final galaxy. A similar convention is
adopted in Heavens and Peacock (1988).

A density peak is characterised by the conditions of zero gradient,
$\nabla\delta={\bf 0}$, and negative definite mass tensor
$\p_{\al}\p_{\beta}\de$. The location of the peak is assumed to
coincide with the origin $\bfq={\bf 0}$. In the vicinity of the
maximum, and in the coordinate system oriented along the principal
axes of the mass tensor $\p_{\al}\p_{\beta}\de$, the density field can be
approximated by (Bardeen \etal 1986) $
\de(\bfq)=\de({\bf 0})-\f{1}{2}\sum_{\al}\lam_{\al}q^2_{\al},
$ where $\lam_{\al}$, $\al=1,2,3$ are the eigenvalues of the tensor
$-\p_{\al}\p_{\beta}\de$. Since the peak corresponds to a maximum, all
$\lam_{\al}$ are positive. Furthermore, it can be assumed that, e.g.,
$\lam_1\geq\lam_2\geq\lam_3>0$. Expressing the height of the peak in
units of $\s_0$, $\de({\bf 0})\equiv \nu\s_0$, the surface boundary of
$\Gamma$ may now be written as $\sum_{\al}A_{\al}^{-2}q_{\al}^2=1$,
where the quantities $A_{\al}$ are the principal semi-axes of the
ellipsoidal isodensity surface $\de_c=0$:
$A^2_{\al}=2\nu\s_0/\lam_{\al}$. Note that this last relation holds
only if $\nu\geq 0$: therefore, all those objects corresponding to
local maxima located in regions of below average mean density (void)
are not taken into account. Fortunately, such objects will probably
fail to form a prominent bound object anyway.

At this point, noting that the inertia tensor $\calJ_{\al\beta}$ is
diagonal in the mass tensor eigenframe, and that
$\Gamma=\f{4\pi}{3}A_1A_2A_3$, one has (see Appendix B for technical
details)
\be
\calL\equiv\sqrt{\lan\bfL^2\ran_{\psi}}=\ell\,\calL_{\ast}\;,
\label{eq:Ltor}
\ee
where we have defined
\be
\calL_{\ast}\equiv a^2\dot{D}\,\eta_0\,\s_0\,R^5_{\ast}\;,
\label{eq:Lref}
\ee
and
\be
\ell\equiv\f{96\,\pi}{\sqrt{15^3}}\,(1-\gamma^2)^{1/2}\,\left(\f{\nu}{\gamma x}\right)^{5/2}\,
\f{\calA(e, p)^{1/2}}{\calB(e, p)^{3/2}}\;.
\label{eq:lsmall}
\ee
The meaning of the peak shape parameters, whose distribution
$\calP(\nu, x, e, p)$ is given in Appendix B, is the following. The
parameter $x$ is an indicator of the \lq sharpness\rq~ of the peak,
sharper peaks have higher $x$; the parameters $e$ and $p$ characterize
the asymmetry of the isodensity profile: $
e\equiv(\lam_1-\lam_3)/2(\lam_1+\lam_2+\lam_3)\,$ and $
p\equiv(\lam_1-2\lam_2+\lam_3)/2(\lam_1+\lam_2+\lam_3)\,$. More in
detail, the parameter $e(\geq 0)$ measures the $ellipticity$ of the
matter distribution in the plane $(q_1, q_3)$, while $p$ determines
the $oblateness$ $(0\leq p \leq e)$ or the $prolateness$ $(-e\leq
p\leq 0)$ of the triaxial ellipsoid. If $e=0$, then $p=0$, and the
ellipsoid is a sphere. The functions $\calA(e, p)$ and $\calB(e, p)$,
which are polynomials in $e$ and $p$, are given explicitly in Appendix
B. For a sphere, $\calA(0, 0)=0$ and $\calB(0,0)=1$.

The meaning of the spectral parameters $\gamma$ and $R_{\ast}$ instead
is the following. The parameter $\gamma$ is a measure of the width of
the power spectrum: if the spectrum is a delta function, then
$\gamma=1$; on the contrary, if $p^7 P_{\psi}(p)$ is constant over a
wide range of $p$, then $\gamma \ll 1$; the dependence of $\gamma$ on
scale for a Cold Dark Matter (CDM) power spectrum is displayed in
Fig.~1 of Bardeen et al. (1986): typically, $\gamma\approx 0.62$ on
galactic scale. The parameter $R_{\ast}$ is a measure of the coherence
scale of the density field or, in other words, it gives an indication
of the smallest wavelength in the power spectrum: $R_{\ast}$ is
related to the smoothing length $R$. For a Gaussian filter and a
power--law spectrum, $R=\sqrt{(n+5)/6\,}\,R_{\ast}$, where $n$ is the
spectral index. To select masses on galactic scale, one typically
takes $R\approx 0.5$ $h^{-1}$ Mpc (for further details see Bardeen et
al. 1986; Appendix B of this paper).

The rms angular momentum $\calL$ in equation~(\ref{eq:Ltor}) has been
written as a product of two factors. The first factor
$\calL_{\ast}=\calL_{\ast}(t)$ contains the temporal dependence and
can be thought of as the angular momentum unit. On the other hand, the
second factor $\ell=\ell(\nu,x,e,p;\gamma)$ is dimensionless and
depends explicitly on the peak shape parameters; in addition, $\ell$
depends on the power spectrum too, through $\gamma$\footnote{Actually,
the factor $(1-\gamma^2)^{1/2}$ in equation~(21) cannot be obtained
directly by specializing equation~(16) to the case of Gaussian peaks,
because it is related to the correlation between inertia tensor
$\calJ$ and the random field $\psi$, which has been neglected in
equation~(13) -- see Appendix B for a self-consistent
derivation. Bearing this in mind, we will continue using the notation
$\lan\,\cdot\,\ran_\psi$, but write the factor $(1-\gamma^2)^{1/2}$
explicitly in the equations that follow.}.  The unit $\calL_{\ast}$
and the shape term $\ell$ are the \lq equivalent\rq~ of $J_{ref}(t)$
and $j_e$ in equations~(10) and (13) in Heavens \& Peacock (1988),
respectively. (Specifically, $\calL_{\ast}$ is identical to
$J_{ref}(t)$ for $f=0$, but $\ell$ is averaged over the constrained
field $\psi$ whereas $j_e$ corresponds to one single realisation of
$\psi\,$, $\ell = \lan j_e^2\ran_\psi^{1/2}$.)

The result in equation~(\ref{eq:Ltor}) is quite simple as a
consequence of our factoring out of the dependence of angular momentum
on the underlying random field $\psi$ on the one hand, and on the peak
shape parameters $\nu,x,e,p$ on the other hand. This is already
manifest in the original equations~(\ref{eq:L2ensem}) and
(\ref{eq:Lensem}) and allows a simplification of the calculation of
the probability distribution of $\ell$, since $\calP(\nu,x,e,p)$ and
the second moment of the off-diagonal shear components is all we need
to know. Heavens \& Peacock (1988) calculated the probability
distribution function of the modulus $L=|\bfL|$ of the peak's angular
momentum. Since $L$ depends on the shear components, they needed to
compute the {\it joint} distribution of shape parameters $\nu$ and
$\lambda_\alpha$ (i.e. $\nu$, $x$, $e$ and $p$), and shear field
$\calE_{\alpha \beta}$; the distribution of early torques on a peak of
given height $\nu$ then leaves a four dimensional numerical
integration. As we will show in Section~3.2, our averaging over the
constrained realisations of the gravitational potential narrows only
slightly the probability distribution of the angular
momentum (see Fig.~7 below).

In the next section, we will compute several distribution functions
for different spin parameters: for this purpose it is useful to express
the mass $M=\eta_0\,\Gamma$ enclosed by the isodensity ellipsoid in
terms of the peak shape parameters:
\be
M\equiv  \mm\, M_{\ast}=
\sqrt{8\,}\left(\f{\nu}{\gamma x}\right)^{3/2}\,
\calB(e, p)^{-1/2}\,M_{\ast}\;,
\label{eq:mass}
\ee
where the reference mass is
$M_{\ast}\equiv\f{4\pi}{3}\,\eta_0\,R_{\ast}^3$.  We stress that the
mass of the collapsing object is estimated as the mass within the
volume $\Gamma$: the filtering procedure, which defines the mass unit
$M_{\ast}$, has to be thought of as corresponding to some genuine
physical filter (more than an artificial one) which prevents the
coupling between the non-linear (small-scale) and the linear
(large-scale) modes in the power spectrum of density fluctuations. The
term $\mm(\nu, x, e, p;\gamma)$ describes the deviation (reduction or
amplification) of the peak mass with respect to the mass unit
$M_{\ast}$ in terms of the shape parameters. We now set the smoothing
scale $R_{\ast}= 0.23\,(0.62/\gamma)^{1/2}\,h^{-1}$ Mpc in order to
have $M_{\ast}=1.45\,(0.62/\gamma)^{3/2}\,10^{10}\,h^{-1}\,M_{\odot}$.
This ensures that mean values of $M$ in equation~(\ref{eq:mass})
correspond to physical masses $\approx 10^{11}M_\odot$ typical of
spiral galaxies and so allows comparison of our angular momentum
distributions with galaxy data. To compare against e.g. cluster data,
one has to choose an appropriate filtering scale $R_\ast$.

It is well known that for individual objects the mass given by
(\ref{eq:mass}) is a rather poor estimate of the mass of the final
collapsed object, as , e.g., shown by numerical simulations. Indeed,
this holds true even for the more sophisticated version of the peak
formalism like the excursion set method (see the review in White
1994). Even so, the mass-function predicted by the latter method, the
Press-Schechter function, is remarkably close to the measured
mass-function. In the same spirit, we suggest that the statistical
predictions which follow from our adopted procedure are more reliable
than the predictions for individual objects.

\section{Angular momentum probability distribution functions}
In this section we compute the probability distribution of three spin
parameters $\ell_\beta\equiv \ell/m^\beta$:
($i$)~$\beta=0$; ($ii)$~$\beta=1$, the specific angular momentum, and
($iii$)~$\beta=5/3$, which is suggested by the dimensional dependence
of angular momentum on mass (see below).

The probability distribution of $\ell_\beta$ can be computed combining
the relation $\ell_\beta=\ell_\beta(\alpha_i)$ with the probability
distribution $\calP(\alpha_k)$, where the $\alpha_k$ ($k=1,\ldots, 4$)
denote the peak shape parameters. For $\calP(\alpha_k)$, we will use
the distribution function given by equation~(\ref{eq:PBard}),
appropriate for Gaussian peaks.

A useful statistic is the conditional probability distribution for
e.g. $\ell_\beta$, given a specific value of one of the parameters
$\alpha_k$, e.g. given the height of the peak. Such a conditional
probability distribution $P(\ell_\beta|\alpha_1)\ d\ell_\beta$ is
formally computed from
\be \calP(\ell_\beta|\alpha_1)\ d\ell_\beta= C_1\, \left(\int d\alpha_3\int
d\alpha_4\, \calP(\alpha_k)\ \left|{\partial \ell_\beta\over \partial
\alpha_2}\right|^{-1} \right) d\ell_\beta\;,
\label{eq:Pc}
\ee
where $C_1$ is a normalization constant such that $\int
\calP(\ell_\beta|\alpha_1)\ d\ell_\beta=1$. The formal
expression~(\ref{eq:Pc}) is general, but we will usually specialize to
the case $\alpha_1=\nu$ and in Appendix C we will show how to compute
the distributions for the different cases $\beta=0,1,5/3$.

The same procedure can be applied to the joint probability
distribution of spin $\ell_\beta$ and mass $m$, $\calP(m,\ell_\beta)\
dm\ d\ell_\beta$:
\be
\calP (m,\ell_\beta)\ dm\ d\ell_\beta = \left(\int d\alpha_3\int
d\alpha_4\, \calP(\alpha_k) \left|{\partial (m,\ell_\beta)\over
\partial (\alpha_1,\alpha_2)}\right|^{-1} \right)\ dm\ d\ell_\beta\;,
\label{eq:N_pk}
\ee
where $m(\alpha_k)$ is the mass of the peak in units of $M_\ast$.
Again, we will show in Appendix C how to specialize the parameters
$\alpha_k$ for the different values of $\beta$.

\subsection{Linear growth era}
During the linear growth era, before the peak decouples from the
expansion of the Universe, angular momentum $\ell$ and peak mass $m$
are related to the peak properties through equations~(\ref{eq:lsmall})
and (\ref{eq:mass}) respectively. Noting that $\ell$ and $m$ 
depend on $\nu$ and $x$ only through the combination $y\equiv \nu/x$, we use
equation~(\ref{eq:N_pk}) to obtain:
$$
\calP(m,\ell_\beta;\nu_1\leq\nu\leq\nu_2)\ dm\ d\ell_\beta= 
$$
\be
\calP_0\left(\int_{0}^{\infty} dy\ \calW[e(y,m,\ell_\beta),p(y,m,\ell_\beta)]
J^{-1}_{ep}J^{-1}_{\calA\calB}\int_{\nu_1/y}^{\nu_2/y}dx\; x^9\ \exp[-f(y)x^2]
\right)\ dm\ d\ell_\beta\;.
\label{eq:N_mj}
\ee
This result is obtained by doing the transformation
$(e,p)\rightarrow(m,\ell_\beta)$ in two steps. Hence, $J_{ep}$ denotes
the modulus of the determinant of the transformation
$(e,p)\rightarrow(\calA,\calB)$ and $J_{\calA\calB}$ is the modulus of
the determinant of the Jacobian of the transformation
$(\calA,\calB)\rightarrow(m,\ell_\beta)$. The function $f(y)$
appearing in the argument of the exponential follows from
equation~(\ref{eq:PBard}). The integral over the peak sharpness $x$
can be performed analytically, yet the inversion
$(\calA,\calB)\rightarrow(e,p)$ cannot, as it involves finding the
real roots of a polynomial of degree six! Further technical details
can be found in Appendix C.

The last equation is a particular case of equation~(\ref{eq:N_pk})
since $\calP(m,\ell_\beta;\nu_1\leq\nu\leq\nu_2)\ dm\ d\ell_\beta$ is
the number of peaks per unit volume, in an interval $dm\ d\ell_\beta$
round specific values $m$ and $\ell_\beta$, which {\it in addition}
have height $\nu$ between $\nu_1$ and $\nu_2$. This is because we wish
to identify a certain class of objects (e.g., spirals, ellipticals,
clusters) assigning a finite $\nu$-interval rather then with a single
value of $\nu$ (see the discussion in Blumenthal et al. 1984).

\begin{figure}
\setlength{\unitlength}{1cm}
\centering
\begin{picture}(10,12)
\put(0.0,0.0){\includegraphics{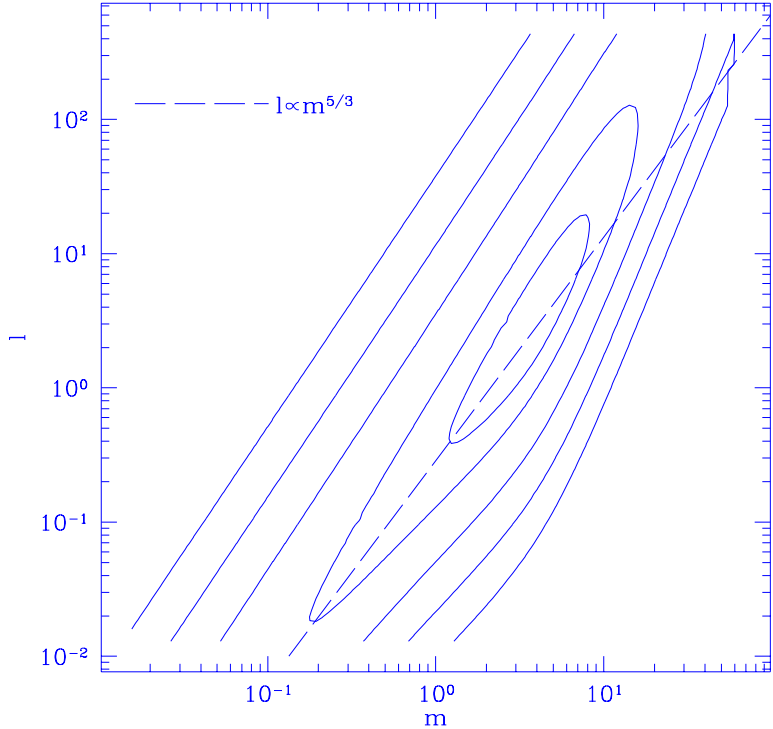}}
\end{picture}
\caption{Joint probability distribution $\calP(m,\ell)$ for CDM
$\gamma=0.62$. Equiprobability contours are drawn at
$\log[m\,\ell\,\calP(m,\ell)\ln(10.)^2]=-9,-7,-5,-3,-2$. The scaling
$\ell\propto m^{5/3}$ is indicated by the dashed line.}
\label{fig:l}
\end{figure}

The joint distribution of $m$ and $\ell$ ($\beta=0,\ \nu_1=0$,
$\nu_2=\infty$) is shown in Fig.~\ref{fig:l} for $\gamma=0.62$,
typical of the CDM model. Note that the most probable value of $\ell$,
given $m$, is strongly correlated with $m$, causing equiprobability
contours to be stretched approximately along $\ell\propto
m^{5/3}$. This dependence is obvious on dimensional grounds, since,
from Eqs.~(\ref{eq:Lref}), $\calL\propto R_\ast^5$ and $M\propto
R_\ast^3$, hence $\calL\propto M^{5/3}$. It follows that
$\calL/M^{5/3}$ is independent of the (rather arbitrary) smoothing
scale $R_\ast$ (see the discussion in Heavens \& Peacock 1988). This
scaling provides an excellent fit to the actual scaling of $\ell$ with
$m$ and is shown in Fig.~\ref{fig:l} for comparison.

\begin{figure}
\setlength{\unitlength}{1cm}
\centering
\begin{picture}(10,10)
\put(0.0,0.0){\includegraphics{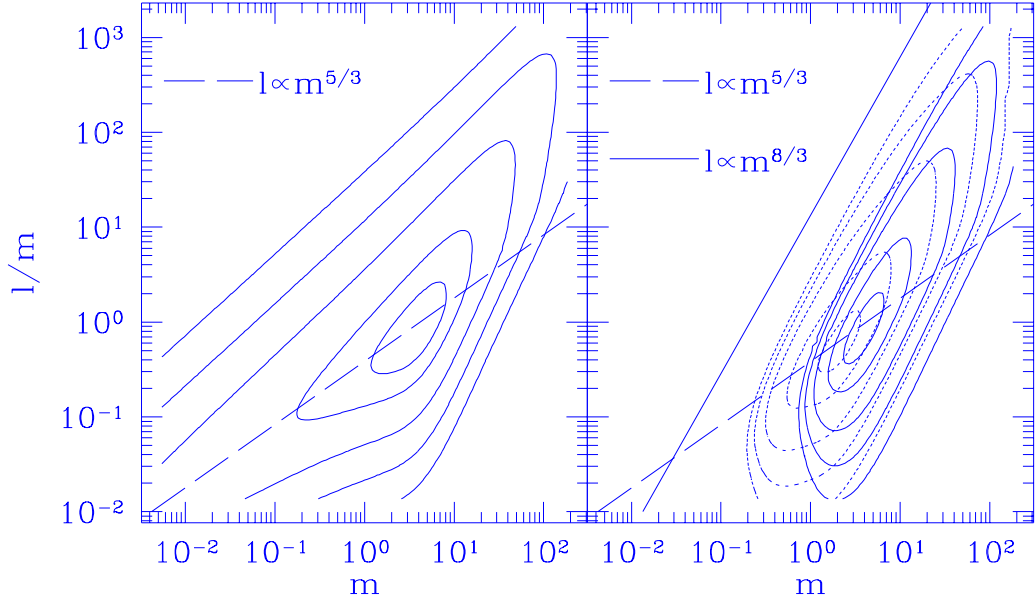}}
\end{picture}
\caption{Same as Fig.~\ref{fig:l}, but for the specific angular
momentum $\ell/m$. Left: equiprobability contours for all objects with
$\nu\geq 0$. Right: dotted equiprobability contours correspond to
objects with $1/2\leq\nu\leq 3/2$ and continuous contours to objects
$3/2\leq\nu\leq5/2$. Scalings $\ell\propto m^{5/3}$ and $l\propto
m^{8/3}$ are indicated.}
\label{fig:lm}
\end{figure}

The joint distribution of $m$ and the specific angular momentum
$\ell/m$ ($\beta=1$) is shown in Fig.~\ref{fig:lm}, again for
$\gamma=0.62$, for the whole range of $\nu$ as well as for restricted
ranges, $1/2\leq\nu\leq3/2$ and $3/2\leq\nu\leq 5/2$. We re-iterate
that our description of angular momentum for protoobjects is limited
to objects with $\nu\geq 0$, hence the {\it total} number of objects
displayed in Fig.~\ref{fig:lm} is only 96\% of the total peak number
density for this value of $\gamma$ i.e., 4\% of peaks have $\nu<0$ for
$\gamma=0.62$. (Note that peaks with $\nu<0$ are {\it maxima} in a
locally underdense region.) The total number of objects with
$1/2\leq\nu\leq3/2$ ($3/2\leq\nu\leq5/2$) in Fig.~\ref{fig:lm} is
40.1\% (37.5\%) of the total number of objects shown. The
equiprobability contours for peaks with the shown restrictions on
$\nu$ display a steeper dependence, $\ell\propto m^{8/3}$, however,
the most probable $\ell$ versus most probable $m$, averaged over
$\nu$, again scales as $\ell_{\rm mp}\propto m_{\rm mp}^{5/3}$. The
figure also shows that objects with the larger value of $\nu$ tend to
be more massive and in addition have a slightly larger $\ell/m$ as
well.

\begin{figure}
\setlength{\unitlength}{1cm}
\centering
\begin{picture}(10,14)
\put(0.0,0.0){\includegraphics{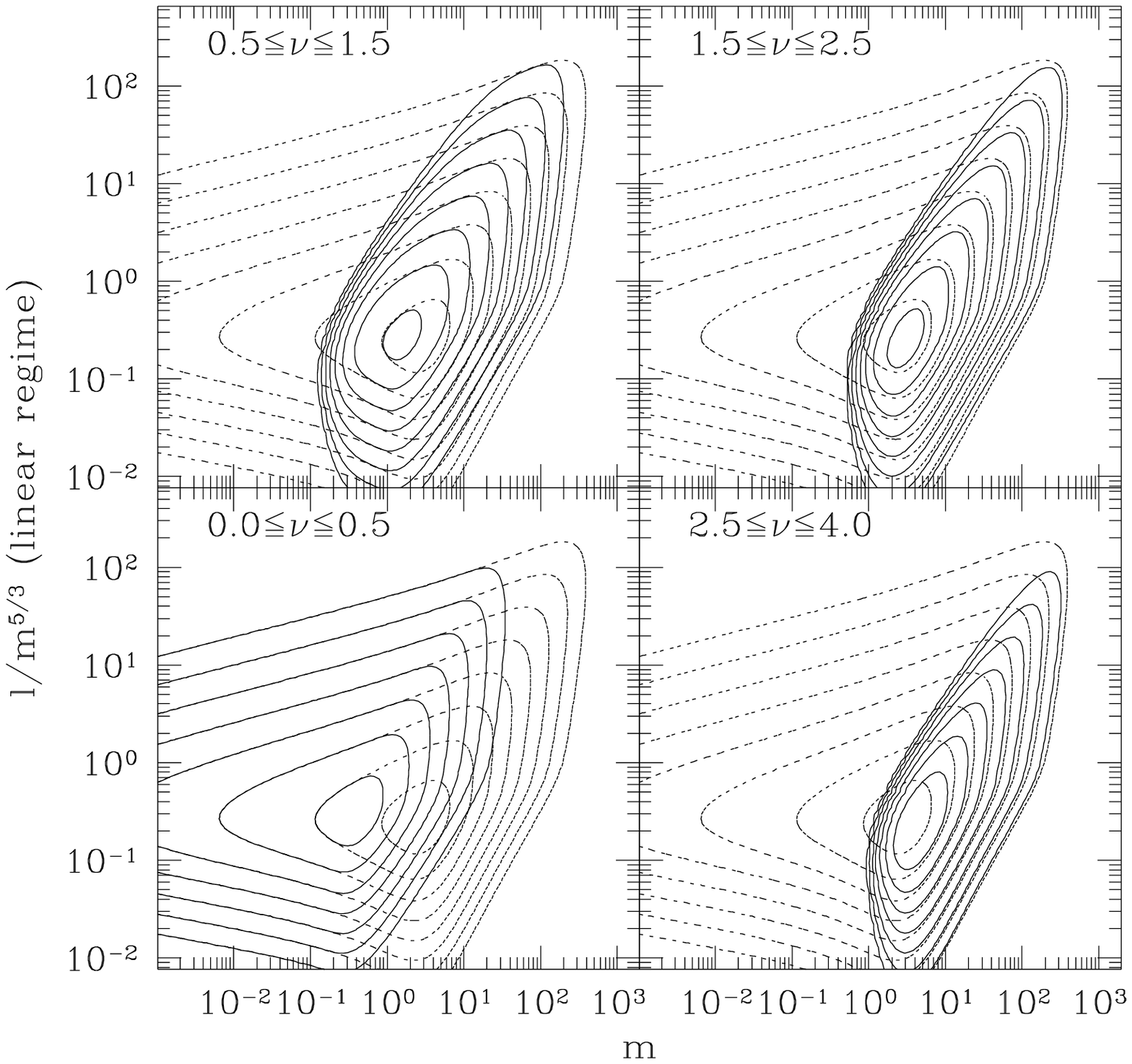}}
\end{picture}
\caption{Joint probability distributions $\calP(m,\ell_{\f{5}{3}}
;\nu_1\leq\nu\leq\nu_2)$ for CDM $\gamma=0.62$. Equiprobability contours
are drawn at
$\log[m\,\ell_{\f{5}{3}}\,\calP(m,\ell_{\f{5}{3}};\nu_1\leq\nu\leq\nu_2)\ln(10.)^2]
=-9,\ldots,-3,-2$. Different panels correspond to different ranges
$[\nu_1,\nu_2]$, as indicated. The dotted contours are identical in
all panels and correspond to the whole range
$[\nu_1=0,\nu_2=\infty)$.}
\label{fig:j53_lin}
\end{figure}

The dependence on peak height $\nu$ is further illustrated in
Fig.~\ref{fig:j53_lin} which shows the joint probability distribution
$\calP(m,\ell_{\f{5}{3}})$, which, in view of the strong dependence
$\ell\propto m^{5/3}$, is the more appropriate quantity to graph. Note
that the most probable value of $l/m^{5/3}$ is very insensitive to
$\nu$, however, the {\it dispersion} round the mean is large.

\begin{figure}
\setlength{\unitlength}{1cm}
\centering
\begin{picture}(10,14)
\put(0.0,0.0){\includegraphics{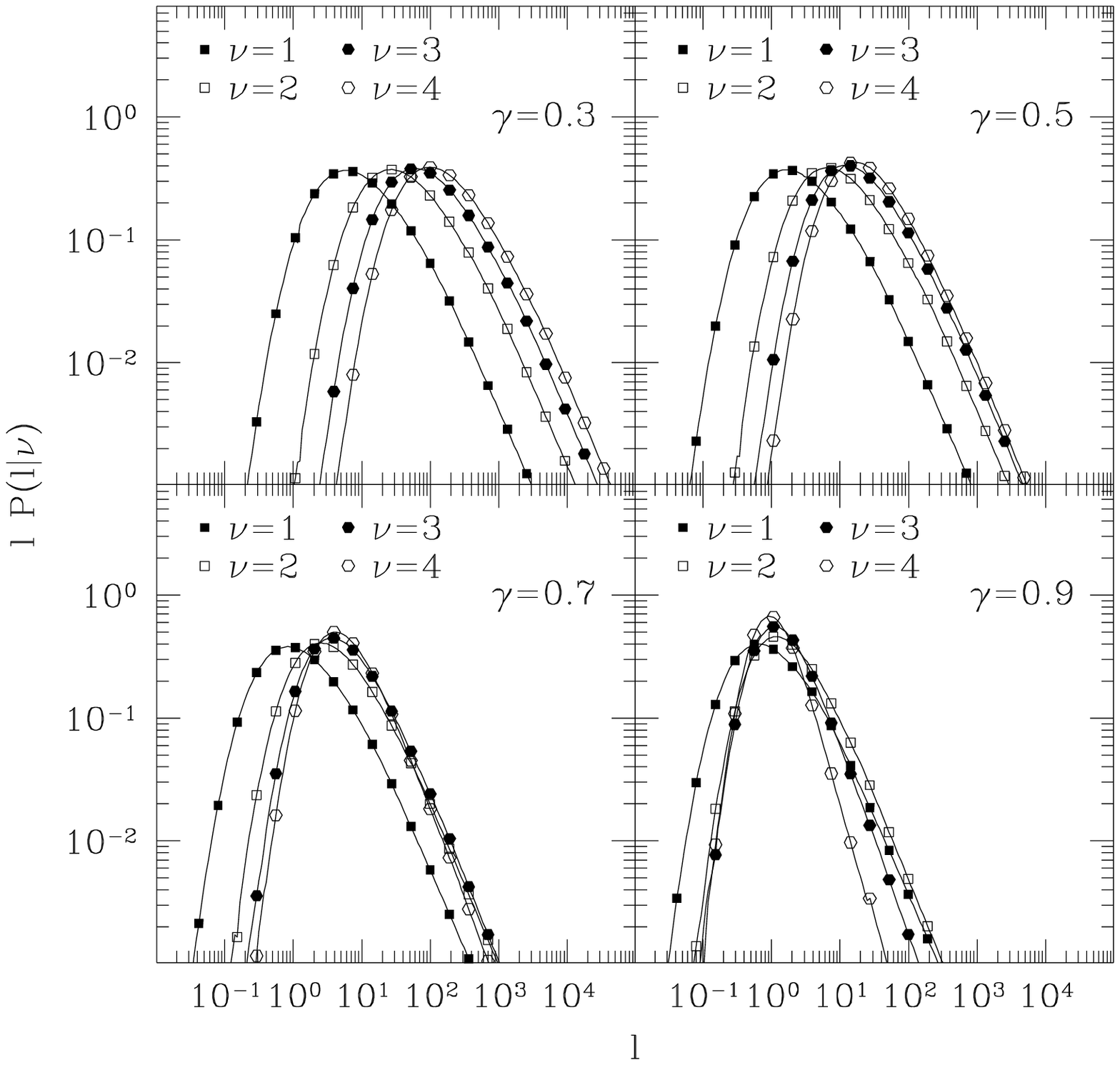}}
\end{picture}
\caption{
Conditional probability for angular momentum $\ell$ during the linear
regime, given the peak height $\nu$, for different values of the
spectral parameter $\gamma$. Different symbols denote different values
of $\nu$, as indicated in the figure. These plots can be compared
against the corresponding ones in Fig.~1 in Heavens \& Peacock (1988)
-- see the discussion in the text.}
\label{fig:Pl}
\end{figure}

\begin{figure}
\setlength{\unitlength}{1cm}
\centering
\begin{picture}(10,14)
\put(0.0,0.0){\includegraphics{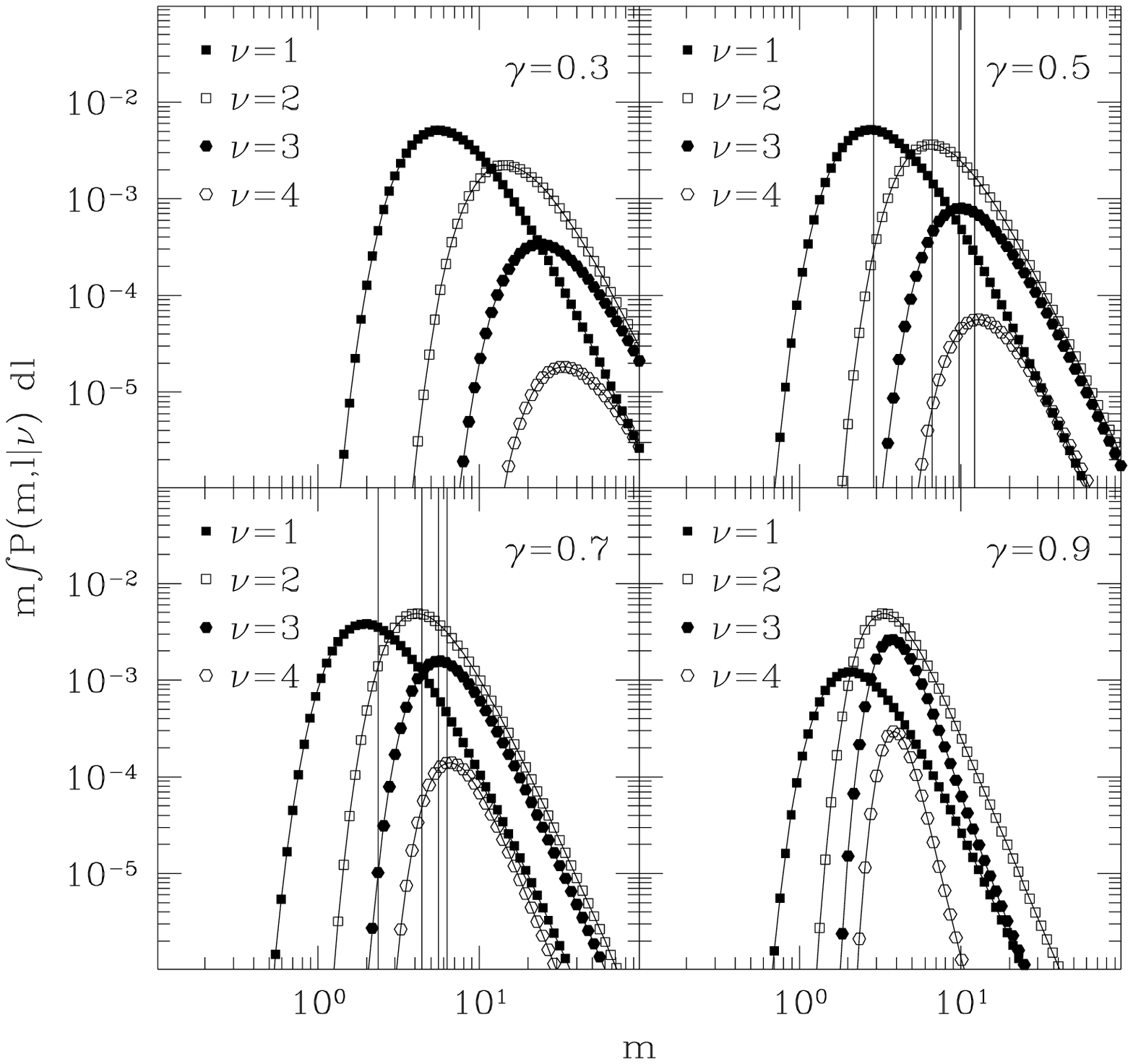}}
\end{picture}
\caption{Distribution of mass for peaks of given height, for several
values of the spectral parameter $\gamma$. Unlike Fig.~\ref{fig:Pl},
mass probabilities shown here are not normalised to one, but reflect
the fact that there are fewer peaks of higher $\nu$. Vertical lines in
panels $\gamma=0.5$ and $\gamma=0.7$ denote the values of the most
probable mass, given $\nu$, obtained from Peacock \& Heavens (1990)
which fit well. Different symbols denote different values of $\nu$,
as indicated in the figure.  }
\label{fig:Pm}
\end{figure}

Conditional probability distributions for $\ell$ and $m$ for given peak height
$\nu$ are shown in Figs.~\ref{fig:Pl} and \ref{fig:Pm}, which also illustrate
the effect of the spectral parameter $\gamma$. Higher $\nu$ peaks have a
higher median $\ell$. This effect is stronger for lower $\gamma$ yet the
distributions are very wide, as was found previously by Heavens \& Peacock
(1988) though for the distribution of the modulus of the angular momentum.

Figure~\ref{fig:Pm} shows the distribution of masses, given $\nu$. In
contrast to Fig.~\ref{fig:Pl}, we have normalised this distribution to
$\int \calP(m|\nu)\ dm\ d\nu=0.016/R^3_\ast$, to illustrate the
different number of objects of given $\nu$, in addition to the
distribution as a function of mass. Again distributions are wide, with
number of objects falling to 1\% of the peak value for masses
differing by factors of five either side of the mean mass.

So far we have restricted ourselves to the growth of the angular
momentum in the linear regime. However, this era of growth is not
generally prone to comparison with observations (but it can be
compared against numerical simulations!) since observed structures
(galaxies, clusters) have already collapsed. Fortunately, objects
presumably acquire most of their angular momentum before they
collapse, since tidal torques are much less effective afterwards
(Peebles 1969, Catelan \& Theuns 1996). However, the non-linear
processes during collapse lead to a redistribution of angular momentum
over the different subunits that will make up the {\it final} object
in a complicated way (White 1984; Barnes \& Efstathiou 1987; Frenk
1987).

We will follow Peebles (1969) in assuming that the linear growth
effectively ceases after maximum expansion of the object and identify
the angular momentum at that time with the \lq final\rq~ angular
momentum. This might be a partial description of reality.

\subsection{Spin distributions at maximum expansion time}
Let us assume that the angular momentum growth described by the
equation~(\ref{eq:L1_White}) stops as the protogalaxy separates from the
overall expansion and collapses back onto itself. A fiducial final angular
momentum $\bfL_f$ (i.e. $\calL_f$) can be calculated as the value of $\bfL$ at
the time $t_{M}$ when, say, $\de(M)=-D(t_M)\nabla^2\psi\equiv
-D_M\nabla^2\psi=1$. From equation~(\ref{eq:L1_White}) and denoting the mass
and radius of the collapsing object by $M$ and $R$ respectively, we get:
\be
L_f \sim a_M^2\dot D_M\nabla^2\psi\, M R^2 = a^2_M\, {\dot D_M\over D_M}\, M
R^2\propto {\dot D_M\over D_M}\, \rho_{bM}^{-2/3}\, M^{5/3} \propto
{\dot D_M\over D_M} \left({\dot a_M\over a_M}\right)^{-4/3} M^{5/3}\;.
\label{eq:L_f1}
\ee
Since in the Einstein-de Sitter universe $a\propto D\propto  t^{2/3}$,
the previous result simplifies to (White 1994)
\be
L_f = L(t_M) \propto M^{5/3}\, t_M^{1/3}\;.
\label{eq:L_f2}
\ee
This expression will be useful when analysing the dimensionless spin parameter
in the next section. Note that, since $t_M$ depends on $M$, the latter equation
does {\it not} imply that $t_M^{1/3}$ is the full temporal dependence of $L_f$
on $t_M$ -- see below.  (In a more general Friedman model, an additional
dependence on the density parameter $\Omega$ appears, which is approximately a
scaling $L_f\propto \Omega^{-0.07}$. Hereafter we will limit ourselves to the
EdS case.)

In the case $M$ refers to the mass of a peak of height $\nu$, the
maximum expansion time $t_M$ depends implicitly on $\nu$, which we
emphasize by writing $t_M\equiv t_{M(\nu)}\equiv t_\nu$. The scaling
of $t_\nu$ with $\nu$ may be recovered extrapolating the linear
spherical model to the maximum expansion time, giving
$t_\nu=\nu^{-3/2}\,t_{\nu=1}$.  From equation~(\ref{eq:Lref}) we find
that the \lq final\rq~ ensemble averaged angular momentum contains an
extra dependence $\calL_f\propto
\nu^{-3/2}$ on peak height in comparison to the linear $\calL$:
\be
\calL_f(t_\nu) = \ell\,\calL_\ast(t_\nu) =
\ell\,\nu^{-3/2}\,\calL_\ast(t_{\nu=1}) \equiv \ell_f \calL_\ast(t_{\nu=1})\;.
\label{eq:L_max_exp}
\ee
The argument goes as follows: start from equation~(20) for the unit of
spin ${\cal L}_\ast$, which we want to compute at the maximum
expansion time (defined by $\nu D(t_\nu)\sigma_0\equiv \nu
D_\nu\sigma_0=1$ for a perturbation of height $\nu$):
\begin{equation}
{\cal L}_\ast (t_\nu) = a_\nu^2 \dot D_\nu\,\eta_0 \sigma_0 R_\ast^5
              = {a_\nu^2 \dot D_\nu\over\nu D_\nu}\,\eta_0 (\nu D_\nu\sigma_0)R_\ast^5
              = {a_\nu^2 \dot D_\nu\over\nu D_\nu}\,\eta_0\,R_\ast^5\,.\nonumber
\end{equation}
Consequently, since $a_\nu^2 \dot D_\nu/D_\nu\propto t_\nu^{1/3}$ in the EdS universe,
\begin{equation}
{\cal L}_\ast(t_\nu) = \nu^{-1}\,t_\nu^{1/3} \eta_0 R_\ast^5\,\propto t_\nu\;.
\end{equation}
Evaluating this for $\nu=1$ gives the
expression~(\ref{eq:L_max_exp})\footnote{The same scaling $L_f\propto
\nu^{-1}\,M^{5/3}\,t_M^{1/3}$ for the modulus of the angular momentum
may be recovered directly from the dimensional analysis in
equation~(\ref{eq:L_f1}), but remembering that ($\alpha\ne\beta$)
$L_f\sim \partial_\alpha\partial_\beta\psi\sim
\nu^{-1}\,\nabla^2\psi$; we recall that the off-diagonal elements of
the shear tensor are {\it independent} of $\nu$ while the trace $\nabla^2\psi$
is proportional to $\nu$.}. Explicitly, $\calL_\ast(t_{\nu=1})$ is calculated
at the time of maximum expansion of a spherical $\nu=1$ overdensity,
$D(t_{\nu=1})\sigma_0=(3/20)(6\pi)^{2/3}$:
\begin{eqnarray}
\calL_\ast(t_{\nu=1}) &=& a^2 \dot
D\,\eta_0\sigma_0R_\ast^5\,|_{t=t_1} = {9 (6\pi)^{2/3}\over 160\pi}
G^{-1} R_\ast^5 \left(a^5 H^3\right)_{t=t_1} = {9 (6\pi)^{2/3}\over
160 \pi} G^{-1} H_0^3 (1+z)_{t=t_1}^{-1/2} R_\ast^5 \nonumber\\ &=&
2.4\times 10^{66} \left({h\over 0.5}\right)^{-2}\,\left({1+z\over
4}\right)^{-1/2}_{t=t_1}\,\left({M_\ast\over 1.45\times
10^{10}h^{-1}M_\odot}\right)^{5/3}\, {\rm
kg~m}^2{\rm~s}^{-1}\nonumber\\ &=&1362\,\left({h\over
0.5}\right)^{-1}\,\left({1+z\over 4}\right)^{-1/2}_{t=t_1} \
M_\ast\,\left({M_\ast\over 1.45\times
10^{10}h^{-1}M_\odot}\right)^{2/3} {\rm km~s}^{-1}\,{\rm kpc}\;.
\label{eq:L_ast}
\end{eqnarray}

Here, $H$ is the Hubble constant, $H_0=100\, h$ km/s/Mpc is
the present day Hubble constant, $z$ is the redshift.
On the other hand, $\ell_f$ is given by
\be
\label{eq:l_max_exp}
\ell_{f} \equiv \nu^{-3/2}\,\ell =
{96\,\pi\over\sqrt{15^3}}\,(1-\gamma^2)^{1/2}
\,\nu^{-3/2}\,({\nu\over\gamma
x})^{5/2}\,{{{\calA}(e,p)}^{1/2}\over {{\calB}(e,p)}^{3/2}}\;.
\ee
The first equality can be compared with equation~(16) in Heavens
\& Peacock (1988), which exhibits the same scaling $\nu^{-3/2}$.

\begin{figure}
\setlength{\unitlength}{1cm}
\centering
\begin{picture}(10,14)
\put(0.0,0.0){\includegraphics{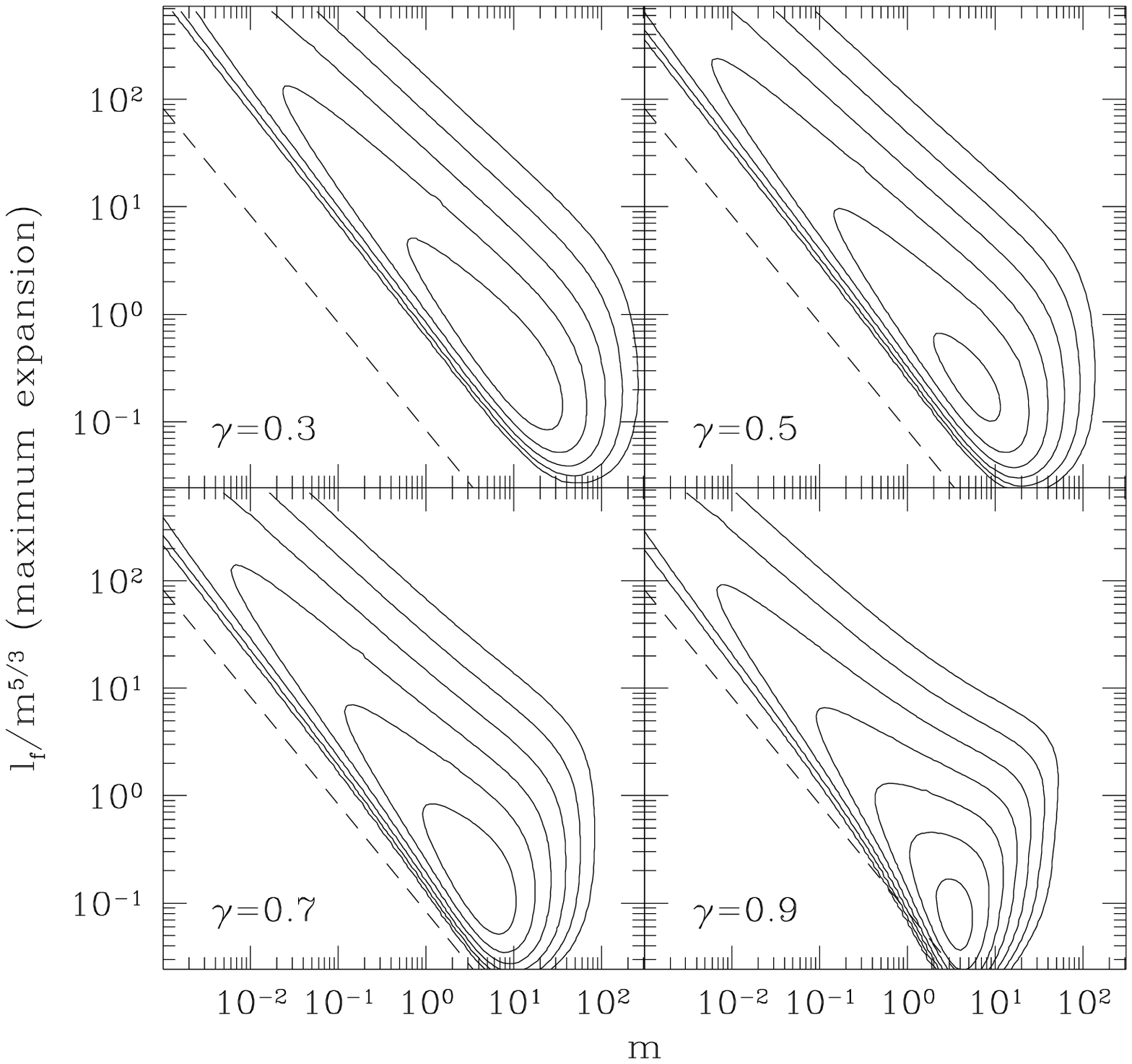}}
\end{picture}
\caption{Joint probability distribution $\calP(m,\ell_{{5\over
3},f})$, where $\ell_{{5\over 3},f}\equiv \ell_f/m^{5/3}$, at maximum
expansion time for different values of the spectral parameter
$\gamma$. Equiprobability contours are drawn at $\log[m\,\ell_{{5\over
3},f}\,\calP(m,\ell_{{5\over3},f})\,\ln(10.)^2]=-9,\ldots,-3,-2$. The
dashed line in each panel is the same and indicates the scaling
$\ell_{{5\over 3},f}\propto m^{-1}$.}
\label{fig:j53_final}
\end{figure}

Figure~{\ref{fig:j53_final}} illustrates the dependence of
$\ell_f/m^{5/3}$ at maximum expansion as a function of $m$ for several
values of the spectral parameter $\gamma$. In all cases,
equiprobability contours are elongated along $\ell_f/m^{5/3}\propto
m^{-1}$, a scaling which follows from the $\ell_f\propto \nu$
dependence read from equation~(\ref{eq:l_max_exp}) and the dependence
$m\propto \nu^{3/2}$, from equation~(\ref{eq:mass}). Note that the
higher $\gamma$ distributions are significantly more peaked round
their maximum.

\begin{figure}
\setlength{\unitlength}{1cm}
\centering
\begin{picture}(10,14)
\put(0.0,0.0){\includegraphics{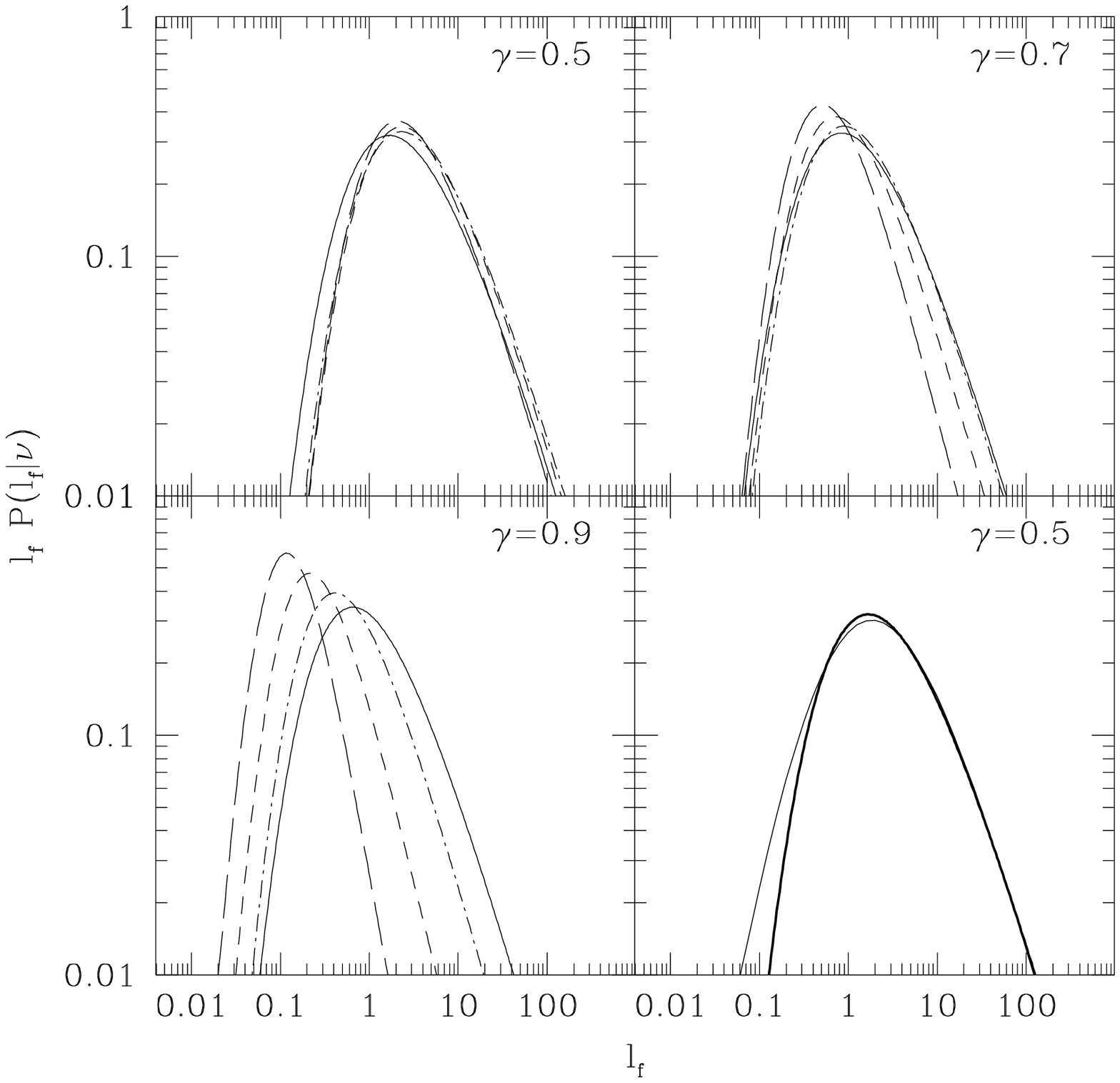}}
\end{picture}
\caption{
Conditional probability distribution for final angular momentum
$\ell_f$ given the peak height $\nu$ for several values of the
spectral parameter $\gamma$. Continuous lines: $\nu=1$, dot-short
dashed: $\nu=2$, short dashed: $\nu=3$, long dashed: $\nu=4$, for
$\gamma=0.5$, 0.7 and 0.9. Lower right panel ($\gamma=0.5$), a
comparison against the Heavens \& Peacock distribution is shown (see
discussion in text). Thick line: distribution for $\nu=1$; thin line:
fit for $\nu=1$ of the distribution shown in Figure~2a of Heavens
\& Peacock (1988),$\calP(\ell_f|\nu=1)=8.2/[(0.89/\ell_f)^{0.25}+(\ell_f/0.89)^{0.47}]^5$.
Note the excellent agreement between these two distributions for
medium and large values of the angular momentum.}
\label{fig:lfinal}
\end{figure}

The conditional probability of final angular momentum $\ell_f$ for
peaks of given height $\nu$ is plotted in Fig.~\ref{fig:lfinal} for
several values of the spectral parameter $\gamma$. Superposed is the
corresponding curve (for $\gamma=0.5$) of Heavens \& Peacock
(1988). The comparison is complicated by the fact that we are plotting
in the same diagram the probability distribution for two different
quantities, essentially $\calP(L_f/\calL_\ast|\nu)$ for the modulus of
the angular momentum and $\calP(\lan
\bfL_f^2\ran^{1/2}_\psi/\calL_\ast|\nu)$ for the rms angular
momentum. This introduces a very small shift between the median values
of $|\bfL_f|$ and $\lan \bfL_f^2\ran^{1/2}_\psi$ which we estimate by
computing the ratio
$\chi\equiv\lan\bfL_f^2\ran^{1/2}/\lan|\bfL_f|\ran$. In the case of a
rotationally symmetric ellipsoid, with $\calJ_{11}=\calJ_{22}$, we
find that $\chi=\lan \calD_{13}^2+\calD_{23}^2\ran^{1/2}/\lan
(\calD_{13}^2+\calD_{23}^2)^{1/2}\,\ran = \sqrt{4/\pi}\approx 1.13$,
using the fact that $\calD_{\alpha\beta}$ is Gaussian distributed with
variance $\sigma_0^2\,(1-\gamma^2)/15$. As Fig.~(\ref{fig:lfinal})
testifies, the distribution of $\calL_f$ is slightly narrower than
that of $L_f$ (because of the shear averaging already performed) but
is still very broad; we stress that the two spin distributions agree
extremely well for medium and large values of the angular momentum.
From this we conclude that the rms angular momentum $\calL_f$ is a
good estimator for $L_f$ and that the loss of information, caused by
averaging over the constrained realisations of the field, appears to
be small. These results are useful when discussing the galaxy
morphology versus rotational properties.

\section{Galaxy morphology and angular momentum}
In this section we compare our predictions against observational data
on both the dimensionless spin parameter $\lambda$ and the specific
angular momentum $L/M$.
\subsection{Dimensionless spin parameter}
A conventional parametrization of galactic angular momenta is the 
dimensionless spin parameter $\lam$
\be
\lam \equiv L \, |E|^{1/2} \,G^{-1/2}\, M^{-5/2}\;,
\label{eq:lambda}
\ee
which is a measure of the ratio between the observed angular velocity
$\w$ and the angular velocity $\w_0$ which would be required to
support the system centrifugally (such as e.g. a rotationally
supported self-gravitating disc): $\lam \approx \w/\w_0\approx
[L/(M\,R^2)]/[(GM/R^3)^{1/2}]$, where $L$ is the modulus of the angular
momentum, $E$ is the total binding energy of the protoobject.

Typically, $\lam$ depends on the galactic morphological type, being as
high as $\lam \approx 0.5$ for spirals and SO galaxies, but only $\lam
\approx 0.05$ for ellipticals, although the dispersion around these
values is large [Efstathiou \& Jones 1979; Fall \& Efstathiou 1980;
Kashlinsky 1982; Fall 1983; Davies \etal 1983, Efstathiou \& Barnes
1984. More in detail, bright giant ellipticals (magnitude $M_B<-19$)
rotate more slowly than faint ellipticals ($M_B>-19$), and the latter
are observed to rotate nearly as rapidly as bulges of spirals and
lenticulars of the same luminosities (Kormendy \& Illingworth 1982;
Illingworth \& Schechter 1982). Also, the flattening of giant
ellipticals is thought to be due to velocity anisotropy rather than to
coherent rotation (Binney 1978)]. We will re-examine below the
question of the correlation between the galaxy morphology and the
amount of angular momentum.

A theoretical estimate of $\lambda$ may be determined from its
ensemble average, $\,\Lam\equiv
\sqrt{\lan\,\lambda^2\ran_{\psi}\,}\,$. From
equation~(\ref{eq:lambda}), one immediately obtains
\be
\Lam \equiv \calL\, |E|^{1/2} \,G^{-1/2}\, M^{-5/2}\;,
\ee
where we have assumed that the total energy $E$ is the $same$ for any
realisation of the potential field $\psi$ (i.e. it is independent of
the deformation tensor $\calD_{\al\beta}$). This is of course a
questionable assumption. It is well known how difficult in general is
the exact calculation of $E$ for a generic enclosing volume
$\Gamma$. Hoffman (1988), expanding the total energy to first order in
$\de$, the gravitational potential to second order in the position and
assuming an ellipsoidal isodensity boundary for $\Gamma$, obtained an
expression for $E$ in terms of the deformation tensor, $E\propto
\calD_{\al\beta}\calJ_{\al\beta}$. Such a dependence on the
deformation tensor of $E$, when consistently taken into account, would
hugely complicate the calculation of $\Lam$.

However, the {\it order of magnitude} of the total energy may be
quantified applying the spherical top-hat model to the evolution of
the overdensity contained in $\Gamma$ (see, e.g., Peebles 1980): we
expect that such a model becomes a better approximation as the height
of the peak increases, since high-$\nu$ peaks tend to be more
spherical.  The spherical model is adopted for the same purpose in
Heavens \& Peacock (1988).

Since the total binding energy of the protogalaxy scales with maximum
expansion time as $E \propto M^{5/3}\,t_{M}^{-2/3}$, then
equation~(\ref{eq:L_f2}) implies that the $final\,$ spin parameter
$\lam_f$ (or, equivalently, $\Lam_f$) does $not$ depend on the
collapse time $t_M$ (actually, this should hold provided that the
protoobject is isolated from the rest of the universe and that
dissipation processes are negligible during the linear evolution).

The resulting expression for the final spin parameter $\Lam_f$ depends
on the peak shape parameters as follows
\be
\Lam_f= {27\,\pi\over 250}\,(1-\gamma^2)^{1/2}\,\nu^{-1}\,\f{\calA(e, p)^{1/2}}{\calB(e,
p)^{2/3}}\;,
\label{eq:Lam_f}
\ee where, in the framework of the spherical model, the expression for
the matter density of the protogalaxy at the maximum expansion time
$t_{M(\nu)}= t_{\nu}$, $\rho(t_{\nu})=(3\pi/4)^2\rho_b(t_{\nu})$, and
the linear prediction of the root mean square density field,
$D(t_{\nu})\s_0=(3/20)(6\pi)^{2/3}\nu^{-1}$, have been used. An
analogous expression for $\Lam_f$, in terms of the quantity
$\ell_{\f{5}{3},f}$ discussed in the previous section, is \be \Lam_f =
\sqrt{{(3/2)^9\over 500}}\, \nu^{1/2}\,\ell_{\f{5}{3},f}\;, \ee which
can be compared against equation~(28) in Heavens \& Peacock
(1988). Note that their equation~(28) for $\lambda_f$ corresponds to a
single realisation of $\psi$ (but recall that their probability
function for $\lambda_f$ does take the shear distribution into
account); however the scaling $\lambda_f\propto\nu^{-1}$ obtained here
is the same: indeed, starting from their definition~(19) of
$j_{{5\over 3}}$, one finds the scaling $j_{{5\over 3}}\propto
\nu^{-3/2}$ from which one recovers the scaling $\lambda_f\propto
\nu^{-1}$ in their equation~(28). The underlying reason for this is
that the ensemble average over the shear does not introduce a $\nu$
dependence.

\begin{figure}
\setlength{\unitlength}{1cm}
\centering
\begin{picture}(10,10)
\put(0.0,0.0){\includegraphics{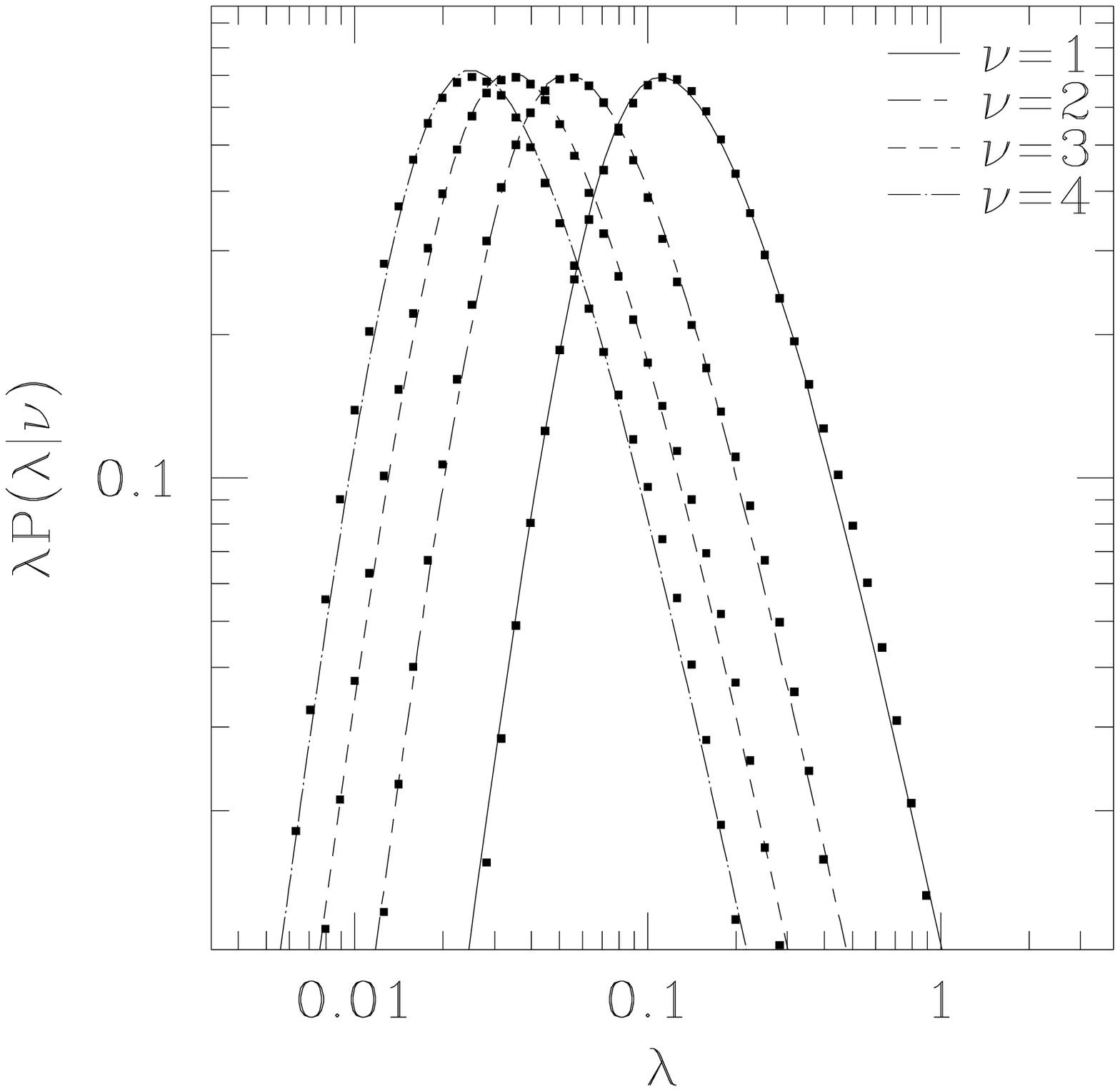}}
\put(0.0,0.0){\includegraphics{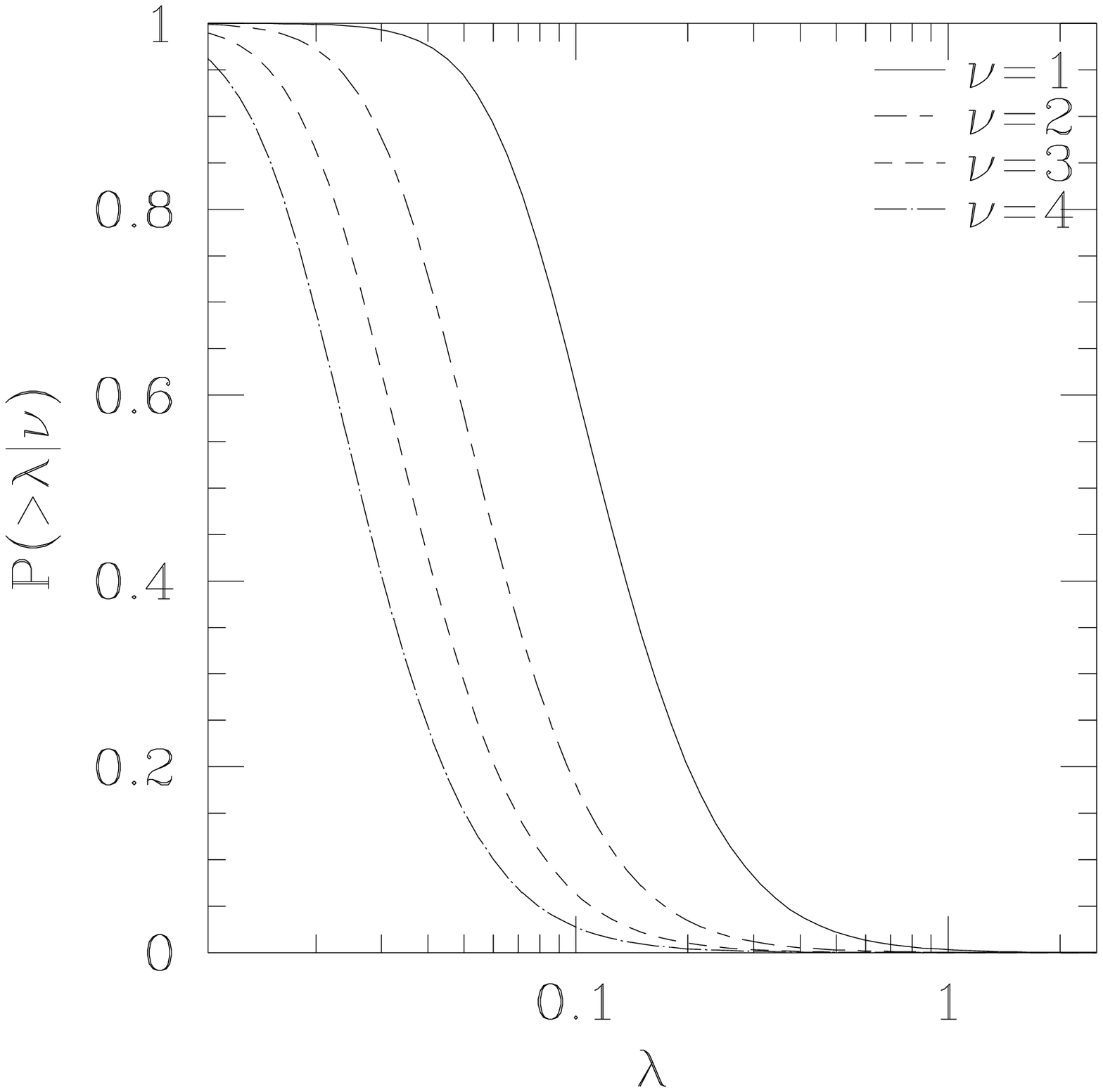}}
\end{picture}
\caption{Left panel: curves denote the conditional probability distributions for
$\lambda$ for different peak heights $\nu$ for a CDM spectrum
($\gamma=0.62$). The squares correspond to the fit given in
equation~(37). Right panel: cumulative distributions for $\lambda$ for
the same $\nu$ and $\gamma$.}
\label{fig:lambda}
\end{figure}

Recently, Steinmetz and Bartelmann (1995) attempted to analyze the $\lambda$
parameter in terms of the peak shape parameters (note that their definitions
of $e$ and $p$ are different from ours). Unfortunately, their statistical
analysis is rather crude since they do not compute the probability
distribution of their $e$ and $p$ and hence are unable to perform the proper
average $\langle\lambda^2\rangle^{1/2}$ (compare their equations~(44) and
~(56)).

The probability distribution function $\calP(\Lam_f|\nu)$, obtained
from equation~(\ref{eq:Lam_f}), is displayed in Fig.\ref{fig:lambda}
and can be compared against Fig.~4 in Heavens \& Peacock (1988),
bearing in mind that our plots are for the rms spin parameter.
Although these authors do not show the case $\gamma=0.62$,
$\gamma=0.7$ is sufficiently close to the CDM value to allow a direct
comparison.

The fit shown in Fig.~\ref{fig:lambda} is
\be
\Lam_f \calP(\Lam_f|\nu) \approx  
\calF_0 \left[1+\alpha\left( {\Lam_f\over
\Lam_0(\nu)}\right)^4\right]\
\exp\left[-{\log^2\left({\Lam_f\over\Lam_0(\nu)}\right)\over
2 \log\Theta}\right]\;,
\label{eq:fit}
\ee
where $\calF_0=0.68$, $\alpha=0.02$, $\Theta=1.12$ and
$\Lam_0(\nu)=0.11\,\nu^{-1.1}\,$. This fits the actual distribution well
for $\nu=1-4$. Note that the median $\overline\Lam_f\approx
\Lam_0(\nu) $ has a slightly steeper dependence on $\nu$ than expected
from equation~(\ref{eq:Lam_f}). (In Fig.~\ref{fig:lambda} we used the
symbol $\lambda$ for $\Lam_f$.)

It can be seen from Fig.\ref{fig:lambda} that higher $\nu$-peaks have
on average lower final angular momentum $\Lam_f$: such a statistical
anti-correlation between $\Lam_f$ (i.e. $\lam_f$) and the peak height
$\nu$ has been first found by Hoffman (1986; 1988) and Heavens \&
Peacock (1988). We also confirm that the $\Lam_f$--distributions are
very broad (factor $\sim $10), and the shift between curves of
different $\lambda$ is small in comparison with the width of the
distribution. Consequently, a given value of $\lambda$ cannot be
unambiguously identified with a specific value of $\nu$.

We stress the fact that the curve corresponding to
$\calP(\Lam_f|\nu=1)$ has the median value $\overline
\Lambda_f=\int\,d\Lam_f\,\Lam_f\,\calP(\Lam_f|\nu=1) = 0.15$. This
brings the value for the $\Lam_f$ for the halos of the progenitors of
spiral galaxies, if assumed to coincide with $\nu=1$ primordial peaks
on galactic scales (Blumenthal \etal 1984), within a factor of two of
the spin parameter of the luminous parts of observed spirals.

It has often been said that energy dissipating cooling processes are
necessary to explain the high spin values of spiral galaxies
(e.g. Efstathiou \& Jones 1979; Fall \& Efstathiou 1980), and
dissipationless $N$-body simulations produce objects with $\lambda\sim
0.08$, too low for spiral galaxies (Barnes \& Efstathiou
1987). However, newer $N$-body simulations which include dissipation
(e.g. Navarro, Frenk \& White 1995) still fail to produce objects with
high $\lambda$. 

If our method for describing tidal torques is sound, then the
importance of cooling processes to boost the value of $\lambda$ from
the values found from dissipationless $N$-body calculations
$\lambda\approx 0.08$ to values $\lambda\approx 0.5$ found for
observed spirals might be partially relaxed. Although we understand
that cooling must be important to explain why the luminous matter of
galaxies is concentrated in the middle of the dark halos, it should
not be invoked as the only cause of the increment of $\lam$ for
spirals with respect to the one for ellipticals, it could be partially
explained on statistical grounds. [Furthermore, we estimate (Catelan
\& Theuns 1996) that the inclusion of mildly non-linear leading-order
corrections to $\bfL$ increases the value of $\lambda$ over that of
the purely linear prediction by a factor $\approx 1.3$.]

Our analysis supports the suggestion that, in hierarchical clustering
scenarios, the higher $\nu$ fluctuations will originate rather smaller
$\lambda$ values, the $\nu\sim 1$ fluctuations on galactic scale being
statistically associated with spiral galaxies and higher $\nu$
fluctuations with ellipticals (Sandage, Freeman \& Stokes 1970;
Kashlinsky 1982; Faber 1982; Blumenthal \etal 1984; Hoffman 1988).

However it is important to recall that our estimate of $\Lam_f$ is
based on an extrapolation of the linear theory and, as discussed in
Barnes \& Efstathiou (1987) and White (1984; 1994), the linear angular
momentum is a relatively poor indicator of the final angular momentum
of a {\it highly} non-linear clump of matter ($\sigma\gg
1$). Non-linear dynamical effects (outward/inward convective motion of
matter etc.), observed to operate in $N$-body simulations, tend to
$reduce$ by a factor $\sim 3$ the final value of $L$ from the
extrapolation of linear theory [although the total scatter about the
mean relation is also a factor of $\sim 3$: see the discussion in
Barnes \& Efstathiou (1987)]. Clearly, one cannot hope to describe
such highly non-linear effects in an exact analytical way.

\subsection{Specific angular momentum}
An obvious advantage of studying the specific angular momentum $L/M$,
rather than the dimensionless parameter $\lambda$, is that the first
does not depend on the binding energy $E$ and therefore is not
affected by uncertainties associated with dissipative processes. In
addition, $L/M$ is not plagued by the problem of how to compute
(analytically) or measure (observationally or from $N$-body
simulations) the energy of the object. Consequently, $L/M$ is a more
robust estimator of the angular momentum of an object and is better
suited to compare different methods of obtaining $L$. (In view of the
strong dependence $L\propto M^{5/3}$, the ratio $L/M^{5/3}$ would be
an even better discriminant.)

\begin{figure}
\setlength{\unitlength}{1cm}
\centering
\begin{picture}(10,10)
\put(0.0,0.0){\includegraphics{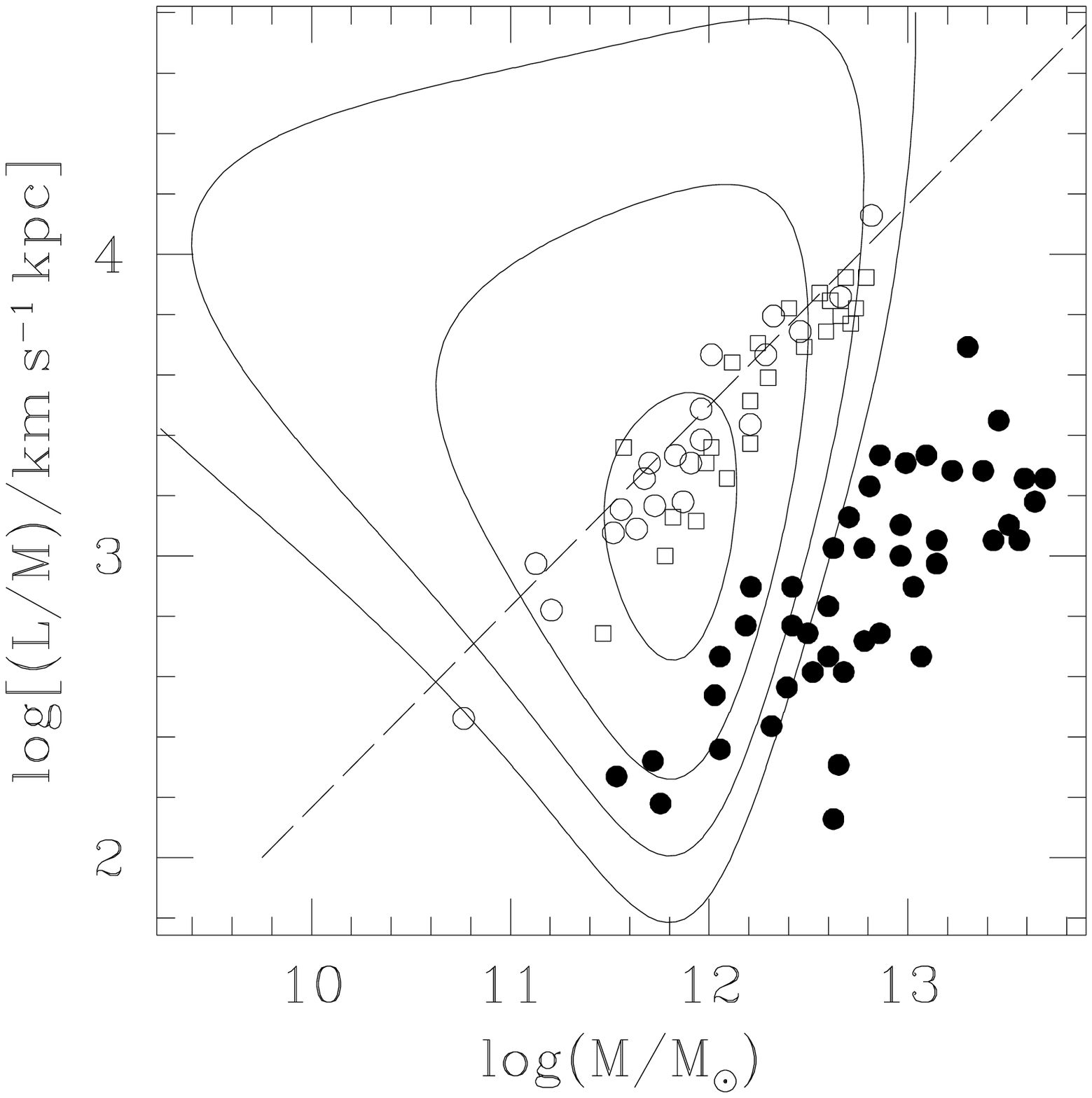}}
\put(0.0,0.0){\includegraphics{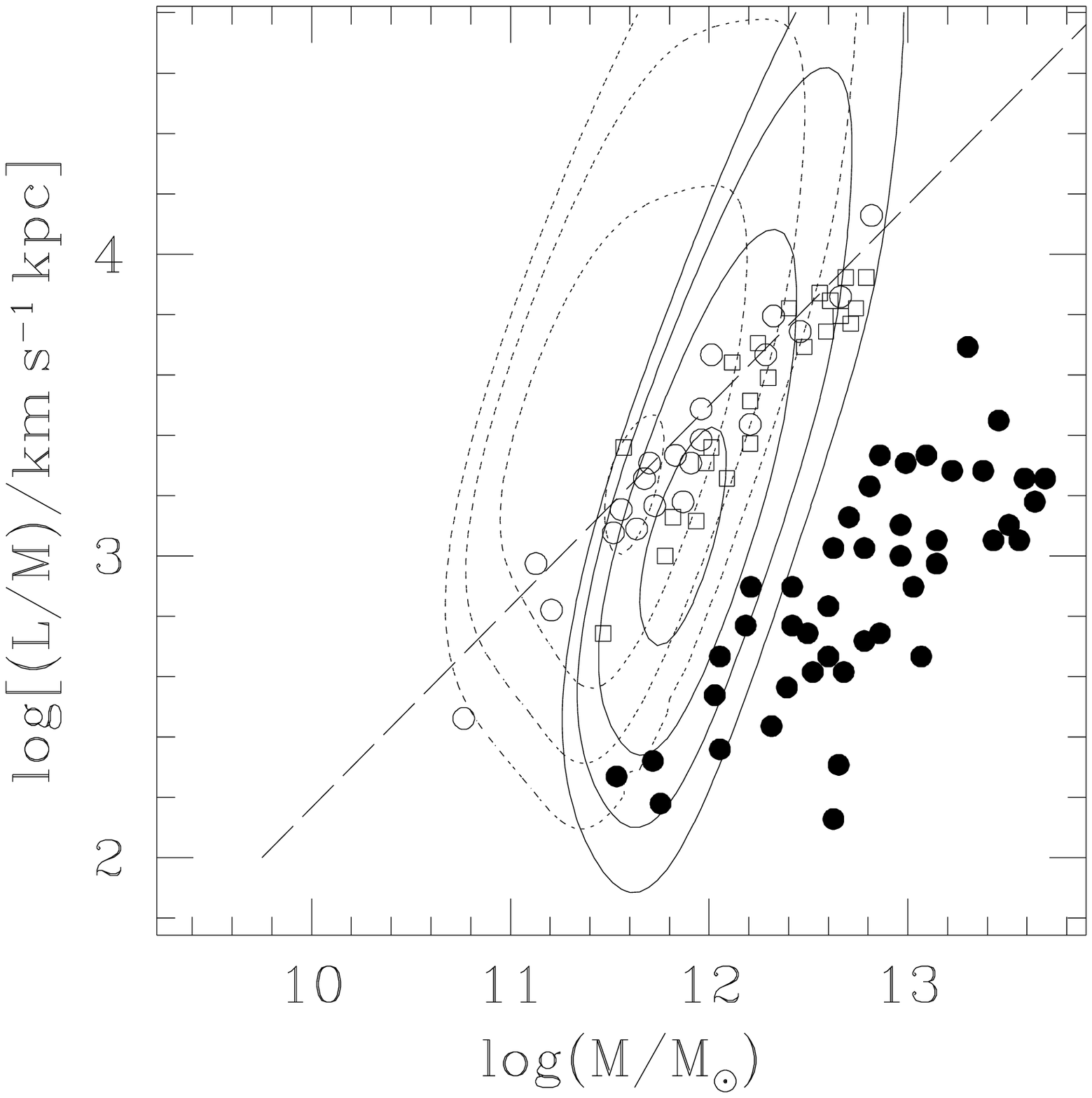}}
\end{picture}
\caption{Symbols: observational data as taken from Fall (1983); $\bullet$:
ellipticals, $\Box $: Sb and $\circ $: Sc. Luminous masses from Fall
are converted to halo masses by multiplying the former by 9.4 (20) for
spirals (ellipticals), keeping $L/M$ constant (see text). Left panel:
equiprobability contours are drawn at
$\log[\calL_f\,\calP(M,\calL_f/M)\,\ln(10.)^2]=-5,\ldots,-2$. Dashed
line indicates the scaling $\calL_f\propto M^{5/3}$, which follows
from the dimensional dependence $\calL_\ast\propto M_\ast^{5/3}$.  The
theoretical equiprobability contours correspond to a single smoothing
scale $R_\ast$, chosen such that $M_\ast=1.74\times 10^{11}\,M_\odot$
for $h=0.5$. Right panel: same as left panel but for peaks with height
$1/2\leq\nu\leq3/2$ (dotted contours) and $3/2\leq\nu\leq5/2$
(continuous contours).}
\label{fig:fall}
\end{figure}

A comparison of our predictions on $L/M$ against observations is made
possible by the data presented in Fig.~1 of Fall (1983), who gives
values of $M$ (luminous) and $L/M$ for a set of elliptical, Sb and Sc
galaxies. This figure shows that the visible parts of the sampled
spiral galaxies have about 6 times as much angular momentum as
elliptical galaxies of the same mass. In Fig.~\ref{fig:fall} we
superimpose the data of Fall (1983) on equiprobability contours of
$\calP(M,\calL_f/M)$ obtained by translating to physical units the
probability distribution $\calP(m,\,\ell_{1,f})$, with $\ell_{1,f}\equiv
\ell_f/m$, calculated in the previous section. We stress that Fall's
masses refer to the luminous parts of the galaxies and so should be
converted to total masses to allow a comparison. Unfortunately, this
procedure is not well defined and we have computed the required
scaling factor as $\sqrt{2}/\lambda$, as suggested in Fall (1983) for
spirals. Taking $\lambda=0.15$ for spirals and $\lambda=0.07$ for
ellipticals, as suggested by our analysis for $\nu\approx 1$ and
$\nu\approx 2$, we find factors 9.4 and 20, respectively. The
smoothing scale is chosen such that the median mass corresponds to the
typical mass of a spiral galaxy. The agreement between the maximum
probability contour and the location of spiral galaxies is rather
remarkable. However, it is clear from the figure that our prediction
of $\calL_f/M$ does not fully accommodate the data of ellipticals which
tend to fall outside high probability contours (but see the discussion
in the last subsection), yet more definite conclusions cannot be drawn
because of the evident smallness of the sample. A larger sample of
data would enhance the fairness of the comparison. The right panel
shows equiprobability contours for objects with a restricted range in
peak height. The location of spiral galaxies is fitted well when
selecting peaks with $\nu\approx 1$. Selecting peaks with $\nu\approx
2$ does tend to shift contours towards the zone occupied by
ellipticals, yet falls short of obtaining good agreement. On the
whole, it does appear that within our description it is not possible
to discriminate between the rotational properties of spirals and
ellipticals {\it solely}~ on the basis of the initial peak height
$\nu$.

It is well known that there is a strong correlation between galaxy
type and environment, in the sense that ellipticals tend to occur in
regions of high galaxy density while spirals tend to avoid such
regions. Hence, interactions between galactic systems, either in the
proto or in the collapsed stages, have an influence on the properties
of the final observed structures. However, even if it were true that
ellipticals form from $\nu\approx 2$ peaks, and although it is well
known that, within the peaks formalism, higher $\nu$-peaks are
naturally clustered more strongly, the likely direct effects of such
enhanced clustering on the {\it tidal torques} on the peaks themselves
are not completely taken into account here. This is best illustrated
recalling the fact that the off-diagonal shear components are {\it
independent} of $\nu$, or, in other words, the (off-diagonal part of
the) shear field does not take into account the effect of the stronger
clustering of higher peaks. So one is possibly missing one of the most
basic ingredients which might generate different types of
galaxies. Consequently, it is likely that elliptical galaxies are not
well described by our formalism. We suggest that, with regard to this
matter, the model might be improved by extending the Taylor expansion
-- which led to equation~(7) -- to higher order.

\begin{figure}
\setlength{\unitlength}{1cm}
\centering
\begin{picture}(10,10)
\put(0.0,0.0){\includegraphics{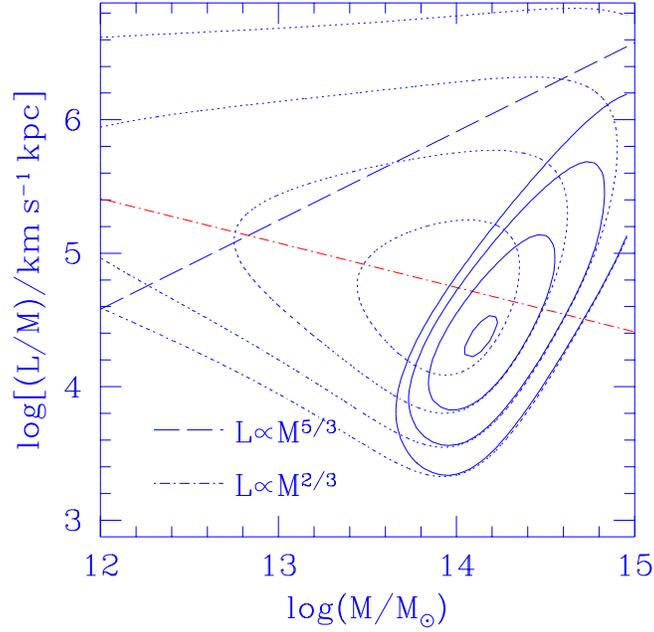}}
\end{picture}
\caption{Equiprobability contours of specific angular momentum versus
mass, using a smoothing scale such that $M_\ast=1.16\times
10^{13}\,M_\odot$ for $h=0.5$ appropriate for rich Abell
clusters. Dotted contours refer to all objects whereas continuous
contours refer to the subset $5/2\leq\nu\leq9/2$. Contour levels are
$-5,\, -4,\, -3,\, -2.2$. These specific angular momenta correspond to
rotational velocities $\approx 5\,$km s$^{-1}$, on cluster scales of 3
Mpc}
\label{fig:clus}
\end{figure}

Our prediction for the dimensional value for $\calL_f/M$ versus $M$ on
the scale of a rich cluster of galaxies is depicted in
Fig.~\ref{fig:clus}. Selecting objects of masses comparable to such
structures leads to typical $\calL_f/M$ values one order of magnitude
larger than for spiral galaxies. Note that the importance of both
non-linear interactions and dissipation are less likely to affect this
prediction than in the comparison with galaxies.

\begin{figure}
\setlength{\unitlength}{1cm}
\centering
\begin{picture}(10,10)
\put(0.0,0.0){\includegraphics{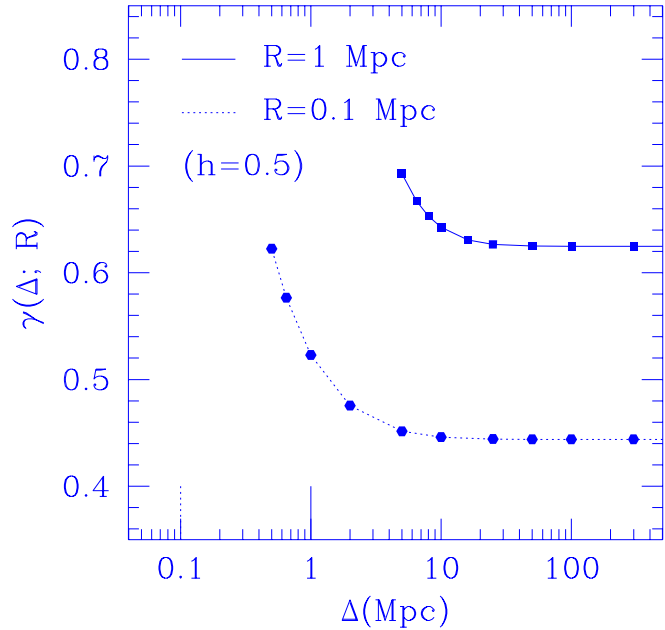}}
\end{picture}
\caption{Effective value  of the spectral parameter $\gamma$ for a CDM
simulation of box size $\Delta$ and resolution $R$. The input CDM
spectrum is from the fit (7.10) in Efstathiou (1989).
}
\label{fig:cdm}
\end{figure}

Our estimate of the specific angular momentum can also be compared
against results from numerical simulations. However, the status of
those simulations seems to be slightly controversial. For instance,
Zurek, Quinn \& Salmon (1988) find values of the specific angular
momentum in agreement with spiral galaxies, but too high for
ellipticals (e.g. their Fig.~2), yet in the calculations of Navarro,
Frenk \& White (1995, Fig.~10), values of $L/M$ are found consistent
with elliptical galaxies but too low for spirals. Note that these
latter authors have included dissipation and find that a substantial
part of the angular momentum of the dissipating gas is lost to the
surrounding halo. However, even their halos have $L/M$ too low to
fall in the region of spiral galaxies, making the discrepancy between
these sets of simulations even worse. Franx, Illingworth \& de Zeeuw
(1991) have voiced the concern that numerical simulations might lack
the dynamic range to be able to describe accurately the problem at
hand. Having a simulation box of size $\Delta$ and Nyquist wavelength
$\sim 1/R$ amounts to changing the effective width of the input power
spectrum and hence has the tendency to change $\gamma$. This effect is
illustrated in Fig.~\ref{fig:cdm}, which shows $\gamma$ for given box
size $\Delta$ and smoothing scale $R$. The bigger box simulation with
$R=1$ Mpc reaches the asymptotic CDM input value
$\gamma=0.62$. Referring to Fig.~\ref{fig:lfinal}, it is clear that
simulations with {\it identical}~ physical spectrum but different
dynamic range may lead to values of the typical angular momentum
differing by factors of a few, especially for rare events.

\subsection{Rotational velocities versus scale}
We can obtain estimates for the rotational velocity from our values of
the angular momentum as a function of mass, for scales corresponding
to galaxies, clusters and superclusters. The angular momentum derived
from equation~(\ref{eq:L_ast}) gives typical values $\calL_f\approx
1.8\times 10^{67}\,{\rm kg~m}^2{\rm ~s}^{-1}$ for a $\nu=1$ peak,
using the median value $\overline{\ell_f}= 5.3$ appropriate for a CDM
$\gamma=0.62$, for $h=0.5$, $z_1=3$ and $M_\ast= 1.84\times
10^{10}h^{-1}M_\odot$ (corresponding to a Milky Way mass $1.1\times
10^{11}\,M_\odot$), well in line with previous estimates for the
Galaxy: Peebles (1969) quotes for the Milky Way a value $2.4\times
10^{67}\,{\rm kg~m}^2{\rm ~s}^{-1}$, comparable to the estimate of
Fall \& Efstathiou (1980). This translates to a circular velocity
$v_R=\calL_f/MR\approx 140\,{\rm km~s}^{-1}$, using $M=1.1\times
10^{11}M_\odot$ and $R=20\,{\rm kpc}$. Applying the same method, we
proceed to compute the typical circular velocity of structures on
larger scales. For a rich cluster of mass $\sim 10^{14}M_\odot$ and
size $R\sim 3{\rm ~Mpc}$, we find $v_R\approx 1.5\times 10^4\,{\rm
km~s}^{-1}{\rm ~kpc}/3\, {\rm Mpc}\approx 5\,{\rm km~s}^{-1}$, where
we used the most likely value of $\calL_f/M$ read from
Fig.~\ref{fig:clus} for $\nu\sim3$. This very low value is clearly
consistent with the lack of observational evidence for cluster
rotation (Efstathiou \& Barnes 1984). For a supercluster of mass
$10^{16} M_\odot$ we use the simple scaling $v_R\sim
5\,\left(\rho_{sc}/\rho_c\right)^{1/3}
\left(M_{sc}/M_c\right)^{1/3}\approx 10\,{\rm km~s}^{-1}$, 
taking the density of the supercluster to be equal to the background
density and the density and mass of a typical galaxy cluster as
previously. Note that these structures are likely to be still in the
linear regime, hence this value must be decreased by the fraction
$t/t_M$ of the age of the Universe in units of the turnaround time of
the supercluster. Furthermore, the spherical approximation is clearly
poor to describe the structures on supercluster scales, hence this
estimate of $v_R$(sc) must be considered rather uncertain.

\section{Summary and conclusions}
In this paper we re-analyzed the problem of the acquisition of angular
momentum by a protoobject, progenitor of a galaxy or a cluster, due to
tidal interactions with the surrounding matter distribution. This
process is of most interest for gravitational instability theories of
galaxy and cluster formation. In section 2, we reviewed the dynamical
description of the spin evolution in the version given by White
(1984), hence the motion of the matter patch of fluid is followed by
applying the Zel'dovich approximation. The expression for the linear
tidal angular momentum $\bfL$ of a protoobject obtained in this
formalism and reported in equation~(7), contains a combination of the
deformation tensor $\calD_{\al\beta}$ and the inertia tensor
$\calJ_{\al\beta}$ of the matter inside the Lagrangian volume
$\Gamma$, which by definition encloses the collapsing
protoobject. Such a combination of the elements of the tensors $\calD$
and $\calJ$ is zero if $\Gamma$ is a spherical volume (or if the
boundary of $\Gamma$ is an equipotential of the gravitational
potential). The temporal evolution of the spin is governed in the
linear regime by the function $a(t)^2\,\dot{D}(t)$, where $a(t)$ is
the scale factor and $D(t)$ is the growth factor of the linear density
perturbations, which is proportional to the cosmic time $t$ in the
Einstein-de Sitter universe (Doroshkevich 1970).

These results may be notably simplified by considering the ensemble
expectation value of the square of the angular momentum
$\lan\bfL^2\ran$. The resulting general (albeit approximate)
expression in equation~(16) is one of the main results of this
paper. It highlights the very simple fact that, in linear regime, the
ensemble average $\lan\bfL^2\ran$ is a function of the first and
second invariant of the inertia tensor $\calJ$ alone and is in
addition proportional to the mass variance $\s_0(R)^2$ on the scale
$R$. The fact that $\lan\bfL^2\ran$ depends only on the invariants of
the inertia tensor and not on the detailed shape of the surface
boundary of $\Gamma$ provides a considerable simplification of the
calculation.

In section~2.3, we specialised the statistic $\lan\bfL^2\ran$ to the
case in which the volume $\Gamma$ is centered on a peak of the
underlying Gaussian density field. The ensemble average is therefore
restricted to those realizations of the density field that are
compatible with the preselected shape of the collapsing object. This
constrained ensemble average corresponds to an unconstrained ensemble
average over the off-diagonal shear (see the extensive discussion in
Appendix~B). The analysis of the tidal torques acting on matter in the
vicinity of local density maxima (peaks) during the linear regime has
been first attempted by Hoffman (1986; 1988) and Heavens \& Peacock
(1988). In particular, Hoffman first analyzed the correlation between
the height $\nu$ of the peak and the amount of angular momentum it
acquires by tidal interactions. Heavens
\& Peacock performed a more sophisticated investigation, taking into
account also the asymmetric shape of the matter peak, described in
terms of the eigenvalues $-\lam_1, -\lam_2, -\lam_3$ of the mass
tensor $\p_\al\p_\beta\de$ (Bardeen et al. 1986). This represents an
important improvement because the strength of the tidal interactions
depends strongly on the shape of the object. Our analysis uses an
approach complementary to the one of Heavens and Peacock and extends
their work.

The resulting expression for $\lan\bfL^2\ran$ appropriate for peaks
has been recast in equation~(\ref{eq:Ltor}) into two factors:
$\calL\equiv\sqrt{\lan\bfL^2\ran}=\ell\,\calL_{\ast}$, where
$\ell=\ell(\nu,x,e,p;\gamma)$ is dimensionless and contains the
dependence on peak shape (height $\nu$, sharpness $x$, ellipticity $e$
and prolateness $p$) and $\calL_{\ast} =
\calL_{\ast}(t;R_\ast)$ is the angular momentum unit which
contains the growth rate and is fixed by the cut-off $R_\ast$ of the
power spectrum. This factorization should be compared with the one in
Heavens \& Peacock (1988) for the modulus $J$ of the angular momentum,
their equation~(9), which corresponds to a single realisation of the
shear field. We then proceed by computing the probability distribution
of several spin variables (e.g. of $\ell$, $\ell/m$, $\ell/m^{5/3}$
versus mass $m$ of the peak; $m$ is the dimensionless mass of the peak
in units of the mass selected by $R_\ast$) by using the probability
distribution of peak shape parameters (Bardeen et al. 1986). This is
an important extension of the work of Heavens \& Peacock (1988), who
analysed the rotational properties of galaxies directly in terms of
spin parameters versus the peak height $\nu$. The latter, however,
does not allow to determine uniquely the mass-scale of the progenitor.
As we argued at the end of section~2.3, we expect our distributions of
angular momentum versus mass to be more reliable than the estimates
for single objects.

In sections~3 and 4, we computed and discussed probability
distributions for $\ell$, $\ell/m$ and $\ell/m^{5/3}$ and the
dimensionless spin parameter $\lambda$, both in linear regime and at
the time of maximum expansion. Our findings can
be summarized as follows:
\begin{itemize}
\item linear values of the spin $\ell$ have a strong dependence on
$m$: $\ell\propto m^{5/3}$, see Fig.~\ref{fig:l}

\item for objects of a restricted range in $\nu$, this dependence
steepens to $\ell\propto m^{8/3}$, see Fig.~\ref{fig:lm}

\item final values (i.e., at maximum expansion time) $\ell_f$ of $\ell$ 
scale shallower with $m$, $\ell_f\propto m^{2/3}$, see
Fig.~\ref{fig:j53_final}

\item median values of $\ell_f$ decrease with $\nu$ for high $\gamma$
(i.e. steep spectra) but are rather insensitive to $\nu$ for low
$\gamma$ (i.e. broad spectra)

\item median values of $\ell_f$ decrease with $\gamma$ for given $\nu$

\item median values for $\lambda$ are $\overline \lambda(\nu=1)=
0.15$ and $\overline\lambda(\nu=2)= 0.07$, comfortably close to the
spin of spiral galaxies but marginally higher than the value typically
quoted for ellipticals

\item dimensional values of $\calL_f/M$ on galactic scales are in
good agreement with measured values for spirals but are too high to
described ellipticals

\item selecting objects on galactic scales with peak height $\nu\sim 1$ gives values
of $\calL_f/M$ typical of spiral galaxies. Selecting objects with
$\nu\sim 2$ moves the position of the most probable $\calL_f/M$ in the
direction of the location of elliptical galaxies in the
$(M,\calL_f/M)$ plane but the shift falls short by a large margin to
give good agreement between values of $\calL_f/M$ for $\nu=2$ peaks
and those observed for ellipticals.

\item typical values of the angular momentum of a spiral galaxy are 
predicted to be $\calL_f\approx 1.8\times 10^{67}\,{\rm kg~m}^2{\rm
~s}^{-1}$ with a corresponding circular velocity $v_{c,{\rm
spiral}}\approx 140\,{\rm km~s}^{-1}$. For a rich cluster, we find a
typical value $v_{c,{\rm clus.}}\approx 5\,{\rm km~s}^{-1}$ while for
a supercluster $v_{c,{\rm sup. clus.}}\le 10\,{\rm km~s}^{-1}.$
\end{itemize}
The reliability of our estimate for the ensemble averaged value
$\Lambda_f\equiv\sqrt{\langle\lambda^2_f\rangle}$ depends on the
extent to which our approximations capture the main features of the
dominant processes. We have made the following assumptions:~(1)
linear perturbation theory, extrapolated to the mildly non-linear
regime (Zel'dovich approximation), can be used to describe the growth
of $\ell$ until the maximum expansion time ;~(2) tidal torques spin up
the matter until the maximum expansion time of the protoobject and are
negligible thereafter;~(3) the mass of the object can be identified
with the mass inside an isodensity ellipsoidal surface around the
peak;~(4) the spherical model can be used to calculate the clump's
binding energy $E$ -- needed to compute $\lambda$;~(5) Gaussian peaks
formalism (Bardeen et al. 1986) can be used to compute probability
distributions for $\ell_\beta$.

We briefly comment on the appropriateness of these approximations.
The use of the Zel'dovich approximation, very powerful in describing
the mildly non-linear evolution of matter before shell-crossing, may
give only a partial description of the evolution of the tidal angular
momentum of the clumps themselves. Certainly, it cannot predict the
very final stages of evolution when clumps merge and interact
non-linearly, which leads to the present-day galactic configurations.
In fact, as shown by the numerical simulations of e.g. Barnes \&
Efstathiou (1987), Frenk (1987), Zurek, Quinn \& Salmon (1988), Navarro
\& Benz (1991) and Navarro \& White (1994), the tidally acquired spin
may change drastically as objects merge or as angular momentum gets
transported between core and halo. The merging history of the
surrounding protohalo is a key factor in the determination of the
morphology of a galaxy and the merging processes of dense clumps are
associated with substantial $loss$ of angular momentum to the
halo. $\calL_f$ -- as obtained from the extrapolation of the linear
theory -- is typically a factor of $\sim$ 3 larger than the final spin
of the non-linear object (Barnes \& Efstathiou 1987; Frenk 1987). At
present, we see little hope of computing theoretically such a
reduction factor, which is due to non-linear interactions as well as
late infall and dissipation processes. Consequently, these processes
constrain the applicability of perturbation theory to the period
before the maximum expansion time (Catelan \& Theuns 1996).

The determination of the surface boundary of the Lagrangian volume
$\Gamma$ is tricky. This boundary determines the mass of the
object. We assumed that it can be described by an isodensity contour,
which is ellipsoidal when the object is centered on a high
peak. However, a non-negligible part of the luminous matter might have
been captured from the very outer regions of the proto-galaxies (Ryden
1988; Quinn \& Binney 1992), where the ellipsoidal model is presumably
a poor description. In addition, the formalism applied here is also
based on Taylor expanding the density and the velocity fields around
the centre of the peak. This expansion breaks down far from the peak's
centre and consequently is unable to accommodate late infall (see the
discussion in Hoffman 1988), a process definitely important during the
late stages of galaxy formation (e.g. Navarro \& White 1994).

In assuming the spherical model to estimate the clump's binding energy
$E$, we have lost the explicit dependence of $E$ on the properties of
the underlying linear gravitational potential field
$\psi$. Consequently, our calculation could be improved by taking this
dependence into account (see Hoffman 1988, section II), but this would
complicate considerably the computation of the ensemble average and is
beyond the scope of the present paper.

Finally, in comparing our predictions against measurements of
luminous parts of galaxies, one has to bear in mind that dissipative
processes, not taken into account here, surely have had a major
influence in shaping those objects.

\section*{Acknowledgements}
James Binney, George Efstathiou and Sabino Matarrese read the original
manuscript of this work. Many and fruitful discussions with Carlos
Frenk, Alan Heavens, Yehuda Hoffman, Bernard Jones, Julio Navarro,
Dennis Sciama, Sergei Shandarin and Rien van de Weygaert are warmly
acknowledged. We owe to the Referee, Alan Heavens, the meticulous
scrutiny of this investigation; the many debates that followed in
Valencia, London and Oxford lead to considerable improvements in both
the presentation and our understanding of this subject. PC and TT
were supported by the EEC Human Capital and Mobility Programme under
contracts CT930328 and CT941463 respectively.

{}

\section*{Appendix A}
In this first appendix, we show explicitly how to derive
equation~(\ref{eq:Lensem}).
The expression in equation~(\ref{eq:dd}) may be concisely written as
\be
\lan\calD_{\beta\s}\calD_{\beta'\s'}\ran_{\psi}=
(\de_{\beta\s}\,\de_{\beta'\s'}+\de_{\beta\beta'}\,\de_{\s\s'}+
\de_{\beta\s'}\,\de_{\beta'\s})\,\Phi\;,
\label{eq:tom}
\ee
where we have defined
\be
\Phi \equiv
\f{4\pi}{15(2\pi)^3}\int_0^{\infty}dp\,p^6\,P_{\psi}(p)\,
\fW(pR)^2={\sigma_0^2\over 15}\;.
\ee
Then, from (\ref{eq:L2ensem})
\begin{eqnarray}
\lan\bfL^2\ran &=&
a^4\dot{D}^2\,\Phi\,\eps_{\al\beta\gamma}\,\eps_{\al\beta'\gamma'}\,
\calJ_{\s\gamma}\,\calJ_{\s'\gamma'}\,
(\de_{\beta\s}\,\de_{\beta'\s'}+\de_{\beta\beta'}\,\de_{\s\s'}+
\de_{\beta\s'}\,\de_{\beta'\s})\nonumber \\
&=&
a^4\dot{D}^2\,\Phi\,
(
\eps_{\al\beta\gamma}\,\eps_{\al\beta\gamma'}\,
\calJ_{\s\gamma'}+
\eps_{\al\s\gamma}\,\eps_{\al\s'\gamma'}\,
\calJ_{\s'\gamma'}+
\eps_{\al\beta\gamma}\,\eps_{\al\s\gamma'}\,
\calJ_{\beta\gamma'})\,\calJ_{\s\gamma}\, \nonumber \\
&=&
a^4\dot{D}^2\,\Phi\,
[
2\,\de_{\gamma\gamma'}\,\calJ_{\s\gamma'}
+
(\de_{\beta\s}\,\de_{\gamma\gamma'}-\de_{\beta\gamma'}\,\de_{\gamma\s})
\,\calJ_{\beta\gamma'}]\,\calJ_{\s\gamma}\,\nonumber \\
&=&
a^4\dot{D}^2\,\Phi\,
[
3\,\calJ_{\al\beta}\,\calJ_{\al\beta}
-(\calJ_{\al\al})^2
]\;.
\label{eq:tom1}
\end{eqnarray}
This scalar expression does not depend on the particular coordinate
system: let us suppose that the axes coincide with the principal
axes of the inertia tensor; then
\be
(\calJ_{\al\beta})\equiv{\rm diag}(\iota_1, \iota_2, \iota_3)\;,
\ee
thus $3\,\calJ_{\al\beta}\,\calJ_{\al\beta}
-(\calJ_{\al\al})^2=2\,(\mu_1^2-3\mu_2)$, where $\mu_1$ and $\mu_2$
are respectively the first and the second invariants of the inertia
tensor, namely $\mu_1\equiv {\rm Tr}(\calJ)=\iota_1+\iota_2+\iota_3$
and $\mu_2\equiv\iota_1\iota_2+\iota_1\iota_3+\iota_2\iota_3$.
Finally, the equation~(\ref{eq:Lensem}) is obtained. We stress that
this result holds for any given Lagrangian volume $\Gamma$ whose
inertia tensor is characterised by the eigenvalues $\iota_\alpha\,.$

\section*{Appendix B}
In this second appendix, we show more in detail how to specialize the 
ensemble average $\lan\bfL^2\ran_{\psi}$ to the case in which the volume 
$\Gamma$ is centered on a peak of the density field: it turns out that 
combination of eigenvalues $\mu_1^2-3\mu_2$ depends on the parameters
characterizing the shape of the peak.

We assume that the isodensity profile $\de_c=0$ is the boundary
surface of the Lagrangian  volume $\Gamma$: such a boundary is 
approximately simply connected and ellipsoidal, at least for sufficiently 
high density peaks (Doroshkevich 1970; Bardeen \etal 1986; Couchman 1987).

A density peak is characterised by the conditions $\nabla\delta={\bf 0}$ 
and the tensor $\p_{\al}\p_{\beta}\de$ has to be negative 
definite: the location of the peak is assumed to coincide with the origin 
$\bfq={\bf 0}$. In the vicinity of the maximum, and in the 
coordinate system oriented along the principal axes of the tensor
$\p_{\al}\p_{\beta}\de$, the density field is approximately described
by the first three terms of the Taylor expansion (Bardeen \etal 1986):
\be
\de(\bfq)=\de({\bf 0})-\f{1}{2}\sum_{\al}\lam_{\al}q^2_{\al}\;,
\ee
where $\lam_{\al}$, $\al=1,2,3$ are the eigenvalues of the
tensor $-\p_{\al}\p_{\beta}\de$: in such a case, each
$\lam_{\al}$ is positive in correspondence of a maximum, and it
can be assumed that, e.g., $\lam_1\geq\lam_2\geq\lam_3>0$.
Expressing the height of the peak in units of $\s_0$, 
$\de({\bf 0})\equiv \nu\s_0$, the surface boundary of $\Gamma$ may 
be written as 
\be
\f{q^2_1}{A^2_1}+\f{q^2_2}{A^2_2}+ \f{q^2_3}{A^2_3}=1\;,
\ee
where the quantities $A_{\al}$ are the principal semi--axes of the ellipsoidal
isodensity surface $\de_c=0$:
\be
A^2_{\al}=\f{2\nu\s_0}{\lam_{\al}}\;.
\ee

At this point, noting that if the tensor $\p_{\al}\p_{\beta}\de$ is diagonal, 
the inertia tensor $\calJ_{\al\beta}$ is diagonal as well,
\be
(\calJ_{\al\beta}) ={\rm diag}(\iota_1, \iota_2, \iota_3) = 
\f{1}{5}\,\eta_0\,\Gamma\,
{\rm diag}(A^2_1, A^2_2, A^2_3)\;,
\ee
it is straightforward to show that the combination
$\mu_1^2-3\mu_2$ reduces to
\be
\mu_1^2-3\mu_2=2^{11}\,3^4\,5^{-2}\,\pi^2\,\eta_0^2\,\left(\f{\nu}{x}\right)^5
\left(\f{\s_0}{\s_2}\right)^5\,\f{\calA(e, p)}{\calB(e, p)^3}\;,
\ee
where we have introduced the polynomials
\be
\calA(e, p)\equiv [p(1+p)]^2+3e^2(1-6p+2p^2+3e^2)\;,
\label{eq:A}
\ee
and 
\be
\calB(e, p)\equiv (1-2p)\,[(1+p)^2-9e^2]\;,
\label{eq:B}
\ee
and the relation $\Gamma=\f{4\pi}{3}A_1A_2A_3$ has to be used.

The parameter
$x\equiv-\s_2^{-1}\nabla^2\de=\s_2^{-1}(\lam_1+\lam_2+\lam_3)$ is an
indicator of the `sharpness' of the peak; the quantity $\s_2$ is
related to the variance of the field $\p_{\al}\p_{\beta}\de$ and it is
an element of a set of spectral parameters weighted by powers of the
wavevector squared, $k^2$,
\be
\s_{h}(R)^2\equiv
\int_0^{\infty}\f{dk\,k^6}{2\pi^2}\,k^{2 h}\,P_{\psi}(k)\fW(kR)^2\;.
\ee
It results that $\lan\, k^2\ran=\s_1^2/\s_0^2$ and $\lan\,
k^4\ran=\s_2^2/\s_0^2$. The parameters $e$ and $p$ characterize the
asymmetry of the isodensity profile:
\be
e\equiv\f{\lam_1-\lam_3}{2(\lam_1+\lam_2+\lam_3)}=
\f{\lam_1-\lam_3}{2\,x\,\s_2}\;,
\ee
\be
p\equiv\f{\lam_1-2\lam_2+\lam_3}{2(\lam_1+\lam_2+\lam_3)}=
\f{\lam_1-2\lam_2+\lam_3}{2\,x\,\s_2}\;;
\ee
also
\be
\lam_1=\f{1}{3}\s_2\,x\,(1+p+3e)\;,
\ee
\be
\lam_2=\f{1}{3}\s_2\,x\,(1-2p)\;,
\ee
\be
\lam_3=\f{1}{3}\s_2\,x\,(1+p-3e)\;.
\ee
More in detail, the parameter $e(\geq 0)$ measures the $ellipticity$
of the matter distribution in the plane $(q_1, q_3)$, while $p$
determines the $oblateness$ $(0\leq p \leq e)$ or the $prolateness$
$(-e\leq p\leq 0)$ of the triaxial ellipsoid.  If $e>0$, then
spheroids with $p=e$ are said to be oblate, while spheroids with
$p=-e$ are said to be prolate; if $e=0$, then $p=0$, the ellipsoid is
a sphere: however, spherical peaks are strongly improbable (see
Bardeen \etal 1986 for more details).

Finally, the ratio $\s_0/\s_2$ in equation~(46) may be written in terms
of two further relevant spectral parameters, $\gamma$ and $R_{\ast}$:
\be
\gamma\equiv\lan x\nu\ran=\f{\s_1^2}{\s_0\,\s_2}\;,
\ee
\be
R_{\ast}\equiv\sqrt{3}\,\f{\s_1}{\s_2}\;.
\ee
The meaning of these statistics is the following: the parameter
$\gamma$ is a measure of the width of the power spectrum, since
\be
\sqrt{\f{\lan\,(k^2-\lan k^2\ran)^2\ran}{\lan\, k^2\ran^2}}=
\sqrt{\f{1}{\gamma^2}-1}
\ee
measures the relative spectral width: if the spectrum is a delta
function, then $\gamma=1$; on the contrary, if $k^7 P_{\psi}(k)$ is
constant over a wide range of $k$, then $\gamma \ll 1$; the dependence
of $\gamma$ on the scale for a Cold Dark Matter power spectrum is
displayed in Fig.~1 of Bardeen et al. (1986): typically, on galactic
scale, $\gamma=0.62$. The parameter $R_{\ast}$ is a measure of the
coherence scale in the field or, in other words, it gives an
indication of the smallest wavelength in the power spectrum (see the
corresponding discussion in the main text).

Inserting the relation $\s_0/\s_2=R_{\ast}^2/3\gamma$ in
equation~(46), and using the general expression~(16) for the ensemble
average $\lan \bfL^2\ran_{\psi}$, and the expression for the volume
$\Gamma$, we obtain
\be
\lan\bfL^2\ran_{\psi} = \f{2^{12}\,3\,\pi^2}{15^3}\,a^4\,\dot{D}^2\,\eta_0^2\,\s_0^2\, 
R^{10}_{\ast}\left(\f{\nu}{\gamma x}\right)^5\,
\f{\calA(e, p)}{\calB(e, p)^3}\;.
\ee
This is an interesting expression: since we $first\,$ calculated the
ensemble average, we have been able to write the expectation value
$\lan\bfL^2\ran_\psi$ as a function of the four peak pattern parameters
$\nu, x, e, p$. Following the suggestion in Heavens \& Peacock (1988),
we can write the previous expression as
\be
\calL\equiv\sqrt{\lan\bfL^2\ran}=\ell\,\calL_{\ast}\;,
\ee
where we have defined
\be
\calL_{\ast}\equiv a^2\dot{D}\,\eta_0\,\s_0\, R^5_{\ast}\;,
\ee
and
\be
\ell\equiv\f{96\,\pi}{\sqrt{15^3}}\left(\f{\nu}{\gamma x}\right)^{5/2}\,
\f{\calA(e, p)^{1/2}}{\calB(e, p)^{3/2}}\;.
\label{eq:lsmallfirst}
\ee
The significance of the quantities $\ell$ and $\calL_{\ast}$
is discussed in the main text.

The probability distribution 
$\calP(\nu, x, e, p)\,d\nu\,dx\,de\,dp$ of the 
peak shape parameters has
been obtained by Bardeen \etal (1986): it reads
\be
\calP(\nu, x, e, p)\,=\,
\calP_0\,x^8\,\calW(e,p)\,
\exp\left[-\f{\nu^2}{2}-\f{5}{2}x^2(3e^2+p^2)-\f{(x-\gamma\nu)^2}{2(1-\gamma^2)}\right]
\;,
\label{eq:PBard}
\ee
in the allowed domain (a triangle) $0\leq e\leq\f{1}{4}\,,\,$ $-e\leq
p\leq e\,$ and $\f{1}{4}\leq e\leq\f{1}{2}\,,\,$ $3e-1\leq p\leq e\,$,
and by definition vanishing outside of the triangle.  The
normalization constant is $\calP_0\equiv 3^2
5^{5/2}/(2\pi)^3R^3_{\ast}\,(1-\gamma^2)^{1/2}$ and the function
$\calW(e,p)\equiv e(e^2-p^2)\,\calB(e, p)$.\\

We stress the fact that the multiplicative factor $(1-\gamma^2)^{1/2}$
appearing in equation~(21), but still missing in
equation~(\ref{eq:lsmallfirst}), can not be obtained by the procedure
outlined so far. The origin of this factor is as follows. The
procedure outlined so far is equivalent to preselecting a volume
$\Gamma$ with a given inertia tensor and {\it independently} assigning
the realisation of $\psi$, hence neglecting the fact that the inertia
tensor is in general statistically correlated with $\psi$. To obtain
consistent results when studying objects with a {\it given} inertia
tensor, the ensemble average of $\bfL^2$ should {\it not} be
calculated over the realizations of the {\it unconstrained} potential
$\psi$ (as we have done so far -- see section~2.2), but only over
those realizations of $\psi$ which are compatible with the preselected
$\Gamma$. In general, this is a hugely difficult calculation to
perform. Fortunately, in the framework of the Gaussian peak model,
this can be done both consistently and easily, in that the angular
momentum is determined by the {\it off-diagonal} components of the
shear field $\calD_{\alpha\beta}$ only, and the latter turn out to be
statistically {\it independent} of the shape of the peak (i.e., of its
inertia tensor). Consequently, as far as the angular momentum is
concerned, to compute the ensemble average over the constrained
potential field, $\lan\cdot\ran_{\psi|\calJ}$, it is enough to compute
the ensemble average over the off-diagonal (unconstrained!)  shear
components, $\lan\cdot\ran_{\calD_{od}}$.

More in detail, the joint distribution of the shape parameters of Gaussian
peaks and the off-diagonal shear components is given by (see equation~(A24) in
Heavens \& Peacock 1988):
\be
\calP(\nu, x, e, p, \calD_{12}, \calD_{13},  \calD_{23} )\,=\,
\calP(\nu, x, e, p)\,\,\calP(\calD_{12}, \calD_{13},  \calD_{23} )\;, 
\ee
where the $\calD$ are Gaussian distributed with zero mean and
variance $\sigma^2_{\calD}=\sigma_0^2\,(1-\gamma^2)/15$:
\be
\calP(\calD_{12}, \calD_{13},  \calD_{23}) =
\left({2\pi\,\sigma^2_{\calD}}\right)^{-3/2}\,
\exp[-(\calD_{12}^2+\calD_{13}^2+\calD_{23}^2)/2\,\sigma^2_{\calD}]\;.
\ee
Note that in equation~(\ref{eq:tom}) we have computed the ensemble
average over the (unconstrained) potential field $\psi$, leading to a
dispersion
\be
\lan\calD_{\beta\s}\calD_{\beta'\s'}\ran_{\psi}=
(\de_{\beta\s}\,\de_{\beta'\s'}+\de_{\beta\beta'}\,\de_{\s\s'}+
\de_{\beta\s'}\,\de_{\beta'\s})\,{\sigma_0^2\over 15}\;,
\ee
whereas computing the ensemble average over the off-diagonal shear (which is
equivalent to computing the average over the constrained potential field) we
now find:
\be
\lan\calD_{\beta\s}\calD_{\beta'\s'}\ran_{\psi|\calJ}=
\lan\calD_{\beta\s}\calD_{\beta'\s'}\ran_{\calD_{od}}=
(\de_{\beta\s}\,\de_{\beta'\s'}+\de_{\beta\beta'}\,\de_{\s\s'}+
\de_{\beta\s'}\,\de_{\beta'\s})\,{(1-\gamma^2)\,\sigma_0^2\over 15}\;.
\ee
Repeating the calculation in Appendix A, but using $\Phi=\sigma_\calD^2$, one
recovers equation~(\ref{eq:lsmall}) instead of equation~(\ref{eq:lsmallfirst}).

Intuitively, the smaller dispersion in the latter case makes sense: the random
field is constrained to have a peak with specified shape, hence the dispersion
of the shear tensor will be smaller than in the first, unconstrained
case. This is the origin of the extra factor $(1-\gamma^2)^{1/2}$ appearing in
equation~(\ref{eq:lsmall}): it properly takes into account the reduction in
rms angular momentum because the reduction of the dispersion in shear due to
the imposed constrained of having a peak with given shape.

A final remark: although the procedure outlined in section~2.2 is only
approximate, it does have some important advantages when one tries to
address fundamental issues related to the tidal torquing of the
progenitors of the present structures. For instance, ($i$) how does
the galactic tidal spin evolve during the non--linear regime? It is
know, in fact, that even if the primordial fluctuations are Gaussian,
the non--linear evolution will ensure that the mass--density
fluctuations become highly non--Gaussian. The simple Gaussian peak
model, for example, cannot be applied during the non--linear regime,
yet the present approximate method can be applied (Catelan \& Theuns
1996); ($ii$) how does the galactic tidal spin evolve during the
non--linear regime starting from intrinsically non--Gaussian initial
conditions? Again, the simple Gaussian peak model cannot be used (work
in progress). The procedure suggested in section~2.2 could give some
insight into these very complicated issues.

\section*{Appendix C}
In this Appendix we give some technical details on how to compute some
probability distribution functions, e.g. the one in
equation~(\ref{eq:N_mj}). We start from the probability distribution
$\calP(\nu,x,e,p)$, given in equation~(\ref{eq:PBard}). The
distribution $P(m,\ell_\beta;\nu_1\leq\nu\leq\nu_2)$ can be obtained
from that by transforming
$(e,p)\rightarrow(m,\ell_\beta)$. Technically, this can be done more
easily in two steps, $(e,p)\rightarrow(\calA,\calB)$ and
$(\calA,\calB)\rightarrow(m,\ell_\beta)$, by using the quantities
$\calA$ and $\calB$ as intermediate variables. To do this we first
invert the relations $\ell_\beta=\ell_\beta(\calA,\calB)$ and
$m=m(\calA,\calB)$, using equations~(\ref{eq:lsmall}) and
~(\ref{eq:mass}), to find:
\begin{eqnarray}
\calA &=& \left({\ell_\beta\over
\ell_{\beta,0}}\right)^2\,\left({m\over m_0}\right)^{2(\beta-3)}\,
\left({\nu\over\gamma x}\right)^4\label{eq:A_ap}\\
\calB &=& \left({m\over m_0}\right)^{-2}\, \left({\nu\over\gamma
x}\right)^3\;,
\label{eq:B_ap}
\end{eqnarray}
where $\ell_{\beta,0}$ and $m_0$ are numerical constants.  One
proceeds by computing the determinant of the Jacobian of the
transformation, $(\calA,\calB)\rightarrow(m,\ell_\beta)$, which is
$J_{\calA\calB} = m\ell_\beta/4\calA\calB$.  On the other hand,
$J_{ep}=216e(e^2-p^2)(1-p^2-3e^3)$ is the determinant of the Jacobian
of the transformation $(e,p)\rightarrow(\calA,\calB)$.  The parameters
$e$ and $p$ occurring in $J_{ep}$ have to be written in terms of
$\calA$ and $\calB$, but this inversion
$(e,p)\rightarrow(\calA,\calB)$ cannot be performed analytically, as
it involves finding the real roots of a polynomial of degree six!
Combining these results, we find for the probability distribution:
\be
\calP(m,\ell_\beta;\nu_1\leq\nu\leq\nu_2)\,dm\,d\ell_\beta = \left( \int_{\nu_1}^{\nu_2} d\nu\;
\int_0^\infty dx\,
\calP[\nu,x,e(\nu,x,m,\ell_\beta),p(\nu,x,m,\ell_\beta)]\,
J^{-1}_{ep}\,J^{-1}_{\calA\calB}\right)\, dm\,d\ell_\beta\;.
\ee
One now observes that the inversion
$(\calA,\calB)\rightarrow(m,\ell_\beta)$ only depends on the ratio $y\equiv
\nu/x$ of the two variables $\nu$ and $x$. Hence, the foregoing
expression can be simplified further in terms of $y$ to give
equation~(25) quoted previously. The function $f(y)$ appearing in the
argument of the exponential in that equation follows from
equation~(\ref{eq:PBard}), $f(y)=y^2/2+5(3e^2+p^2)/2+(1-\gamma
y)^2/2(1-\gamma^2)$. One can now perform the integral over the peak
sharpness $x$ analytically, leaving a one dimensional numerical
integration to obtain $\calP (m,\ell_\beta;\nu_1\leq\nu\leq\nu_2)$.

In some cases (e.g. to compute $\lambda$ or $\ell_f$) there is an
extra $\nu$ dependence which makes that $\calA$ and $\calB$ do not
depend on the ratio $\nu/x$ alone. In the computation of the
distribution $\calP(m,\ell_f/m^\beta)$, we eliminated $\nu$ and $x$ in
favour of $m$ and $\ell_f/m^\beta$, and $\calP$ follows from a two
dimensional integration over $e$ and $p$. The computation of the
$\lambda$ distribution for given $\nu$, on the other hand, can again
be obtained from a one dimensional numerical integration, by
transforming $(e,p)\rightarrow (m,\lambda)$ and noting that the latter
change of variables only depends on $y\equiv m\,x^{3/2}$ and $x$,
which allows one to compute the $x$ integral analytically. We do not
give the latter analytical expressions since they are rather
cumbersome.

\end{document}